\documentclass{aa}

\usepackage{graphicx}
\usepackage{amsmath}
\usepackage{txfonts}
\usepackage{natbib}
\bibpunct{(}{)}{;}{a}{}{,}
\usepackage{hyperref}

\begin{document}

\title{Combining weak and strong cluster lensing: Applications to simulations and MS~2137}
\author{Julian Merten \inst{1} \and Marcello Cacciato \inst{2}  \and Massimo Meneghetti \inst{3,4} \and Claudia Mignone \inst{1,2} \and Matthias Bartelmann \inst{1}}
\institute{Institut f\"ur Theoretische Astrophysik,~Zentrum f\"ur Astronomie der Universit\"a~Heidelberg, Albert-Ueberle-Str.2, 69120~Heidelberg, Germany\\
\email{jmerten@ita.uni-heidelberg.de}
\and
Max-Planck-Institut f\"ur Astronomie, K\"onigstuhl~17, 69117~Heidelberg, Germany
\and
INAF-Osservatorio Astronomico di Bologna, Via Ranzani 1, 40127~Bologna, Italy
\and
INFN-National Institute for Nuclear Physics, Sezione di Bologna, Viale~B.~Pichat~6/2, 40127~Bologna, Italy}

\date{Accepted for publication in A\&A on March~28, 2009.}

\abstract{}
{While weak lensing cannot resolve cluster cores and strong lensing is almost insensitive to density profiles outside the scale radius, combinations of both effects promise to constrain density profiles of galaxy clusters well, and thus to allow testing of the CDM expectation on dark-matter halo density profiles.}
{We develop an algorithm further that we had recently proposed for this purpose. It recovers a lensing potential optimally reproducing observations of both strong and weak-lensing effects by combining high resolution in cluster cores with the larger-scale information from weak lensing. The main extensions concern the accommodation of mild non-linearity in  inner iterations, the progressive increase in resolution in outer iterations, and the introduction of a suitable regularisation term. The linearity of the method is essentially preserved.}
{We demonstrate the success of the algorithm with both idealised and realistic simulated data, showing that the simulated lensing mass distribution and its density profile are well reproduced. We then apply it to weak and strong lensing data of the cluster MS~2137 and obtain a parameter-free solution which is in good qualitative agreement with earlier parametric studies.}{}


\maketitle


\section{Introduction}
\label{Introduction}

The mass distribution in dark-matter halos and the level of substructure in them are among the central predictions of the CDM paradigm for cosmic structure formation. The density profile should asymptotically fall off $\propto r^{-3}$ at large radii $r$ and flatten considerably within a radial scale $r_\mathrm{s}$ \citep{Navarro1996}.The mass distribution should be richly substructured by sublumps of matter with a differential mass function approximated by a power law, $d n/d M\propto M^{\alpha}$ with a slope slighly shallower than $\alpha=-2$ \citep{Madau2008,Springel2008}.

Galaxy clusters should be weakly influenced by baryonic physics, thus their density profiles and mass distributions outside the cooling radius should well reflect those expected for dark matter. Do they? Although tentative answers exist, showing that estimated density profiles do at least not contradict the CDM expectation, accurate constraints are still missing. Due to its insensitivity to the physical state of the matter, gravitational lensing is perhaps the most promising tool for determining matter distributions.

Weak lensing lacks the resolution necessary to constrain the density profile in cluster centres, while strong lensing is confined to the innermost cluster cores. In combination, they may be able to test the CDM predictions on density profiles well.

Several methods have been suggested to combine weak and strong cluster lensing \citep{Bradav2005,Cacciato2006,Diego2007}. Among them is our own algorithm aiming at the lensing potential. It is based on minimising a $\chi^2$ function comparing observed shear measurements with suitable second derivatives of the potential. Expressing the derivatives in terms of finite differences leads to a system of linear equations whose direct inversion yields the solution.

We extend our earlier work in several ways. First, we no longer use the lowest-order approximation in which measured ellipticities estimate the shear, and introduce the reduced shear instead. The non-linearity accommodated in this way can be resolved into an iterative scheme using linear inversion in each step. Second, we wrap the algorithm into an outer iteration loop in which the grid resolution is progressively enhanced. While this step introduces correlations between adjacent pixels that have to be dealt with, it prepares the insertion of the strong-lensing constraints available in cluster cores. Third, we introduce a regularisation term for the two purposes of avoiding overfitting and smoothly joining the strong- and weak-lensing solutions. Finally, to account for the additional computational time we enabled the code to run on parallel machines.

We investigate the performance of our algorithm using two sets of synthetic data, one idealised and one realistic, before we proceed to apply it to the well-known strong-lensing cluster MS~2137, for which we obtain a high-resolution, parameter-free reconstruction.

A brief summary of the lensing notation in Sect.~\ref{lensingformalism} is followed by an outline of the method in Sect.~\ref{outlineofthemethod} and a description of its implementation in Sect.~\ref{implementation}. We present the results in Sect.~\ref{results} and conclude in Sect.~\ref{conclusions}. Details of the algorithm are given in Appendix~\ref{linearisationingridspace}.


\section{Lensing formalism}
\label{lensingformalism}

\subsection{Basic quantities}
\label{basicquantities}

We adopt the standard notation introduced to describe isolated lenses in the thin-lens approximation (e.g.~\citealt{P.1992,Narayan1996,P.2006}). Two-dimensional, projected lensing mass distributions are covered by angular coordinates $\vec{\theta}=(\theta_{1},\theta_{2})$. The lensing potential $\psi(\vec\theta)$, which is the appropriately scaled Newtonian potential projected on the sky, contains all information necessary to describe a single-plane lens. The deflection angle, convergence and shear are derivatives of $\psi(\vec{\theta})$ with respect to $\theta_{1}$ and $\theta_{2}$,
\begin{align}
  \vec{\alpha}&=\nabla\psi \label{defang} \\
  \kappa &= \frac{1}{2} \nabla^{2} \psi =
  \frac{1}{2}\left(\frac{\partial^{2}}{\partial\theta_{1}^{2}}+\frac{\partial^{2}}{\partial\theta_{2}^{2}}\right)\psi=
  \frac{1}{2}(\psi_{,11}+\psi_{,22}) \label{convergence} \\
  \gamma_{1} &= 
  \frac{1}{2}\left(\frac{\partial^{2}}{\partial\theta_{1}^{2}}-\frac{\partial^{2}}{\partial\theta_{2}^{2}}\right)\psi=
  \frac{1}{2}(\psi_{,11}-\psi_{,22}) \label{shear1} \\
  \gamma_{2} &= \frac{\partial^{2}}{\partial\theta_{1}\theta_{2}}\psi=\psi_{,12}\;.
\label{shear2}
\end{align}

\subsection{Lensing by galaxy clusters}
\label{lensingbygalaxyclusters}

We concentrate on lensing by galaxy clusters. Let us first focus on weak lensing, which means $\kappa \ll 1$ (see \citet{Bartelmann2001} for a review). To first order, shape distortions of background galaxies are determined by the Jacobian matrix of the lens mapping,
\begin{equation}
  \mathcal{A}(\vec{\theta})=\left(
    \delta_{ij}-\frac{\partial^{2}\psi(\vec{\theta})}{\partial\theta_{i}\partial\theta_{j}}
  \right)=\begin{pmatrix}
    1-\kappa -\gamma_{1} & -\gamma_{2} \\
    -\gamma_{2} & 1-\kappa +\gamma_{2}
  \end{pmatrix}\;.
\label{jacobian}
\end{equation}
The intrinsic ellipticities of the background galaxies require that their images be averaged to extract the weak-lensing signal from them. We assume that averages over ten or more galaxies are necessary for the uncertainty of an individual ellipticity measurement to fall below 10\% of the signal \citep{Cacciato2006}. The expectation value for the measured ellipticity is then
\begin{equation}
  \left<\varepsilon\right>=\left\{\begin{aligned}
    \frac{Z(z)\gamma}{1-Z(z)\kappa} &\qquad \textrm{for }|g|\leq 1 \\ 
    \frac{1-Z(z)\kappa}{Z(z)\gamma^{*}} &\qquad \textrm{for } |g| > 1
  \end{aligned}\;,\right.
\label{reducedshear}
\end{equation}
where the reduced shear
\begin{equation}
  g(\vec{\theta})\equiv\frac{\gamma(\vec{\theta})}{1-\kappa(\vec{\theta})}
\end{equation}
and the distance weight function
\begin{equation}
  Z(z)\equiv\frac{D_{\infty}D_{\text{ds}}}{D_{\text{d}\infty}D_{\text{s}}}H(z-z_{\text{d}})
\end{equation}
appear. In the last equation, $z_{\text{d}}$ is the redshift of the lens, while $D_{\infty}$ and $D_{\text{d}\infty}$ are the angular-diameter distances between observer and infinity and between lens and infinity, respectively.

While the probes of weak lensing are slightly distorted background galaxies whose signal needs to be treated statistically, strong lensing is based on greater effects. Another difference from weak lensing is that strong lensing only occurs in galaxy clusters near their cores where the lens becomes critical. These regions are typically of the order of 100 kpc in radius. The main observations are

\begin{itemize}

\item multiple images of background sources, which all carry the same spectral information of the source, which enables their unambiguous identification through a spectral or colour analysis, and

\item highly distorted images of background sources like gravitational arcs or arclets, which lie close to critical curves of clusters.

\end{itemize}

Critical curves are closed point sets in the lens plane where the Jacobian becomes singular,
\begin{equation}
  \det\mathcal{A}_{\text{crit}}=\left(1-\kappa_{\text{crit}}\right)^{2}-
  |\gamma_{\text{crit}}|^{2}=0\;.
\end{equation}


\section{Outline of the method}
\label{outlineofthemethod}

Our non-parametric maximum-likelihood reconstruction method aims at recovering the lensing potential $\psi$. The reasons for this choice are that the lensing potential is much smoother than e.g.~the convergence, which renders it much less susceptible to noise, and that both convergence and shear are derivatives of the lensing potential so that no integration is needed to convert one to the other. The method described and applied here develops further and extends those presented in \citet{Bartelmann1996,Seitz1998,Cacciato2006}.

The method takes as input the result of galaxy-shape and strong-lensing measurements, i.e. the two ellipticity parameters per galaxy and the strongly lensed images at their angular positions. As we pointed out before, an ellipticity measurement of a single galaxy image is useless as a weak lensing signal because of the intrinsic source ellipticity. Each data point is thus obtained by averaging over a certain number of background galaxy ellipticities.

We then divide the observed galaxy-cluster field into a grid of $N$ cells, assign an averaged ellipticity to each cell, and thus obtain $N$ data points for our $\chi^{2}$-minimisation. The ensuing reconstruction will strongly depend on the grid resolution. Furthermore, if a number of $M$ grid cells, which we shall call pixels from now on, contains strongly lensed images, we gain $M$ additional constraints for the reconstruction.

Since the weak and strong-lensing constraints are independent of each other, but reflect the same underlying gravitational potential, the overall $\chi^{2}$ becomes the sum of two independent contributions,
\begin{equation}
  \chi^{2}(\psi)=\chi^{2}_{\text{w}}(\psi)+\chi^{2}_{\text{s}}(\psi)\;.
\label{chi2}
\end{equation}
Our method is non-parametric in the sense that it does not assume a parameterised model for the mass or potential distribution. It assigns an initially unknown potential value to each grid point and refines the set of potential values on the grid during the $\chi^{2}$-minimisation. We are thus searching for a discrete representation of the lensing potential, which is optimally capable of reproducing the observed lensing effects.

The reconstruction proceeds by minimising $\chi^{2}$ with respect to the potential values $\psi_l$ at all grid positions $l$,
\begin{equation}
  \frac{\partial\chi^{2}(\psi)}{\partial\psi_{l}}=
  \frac{\partial\chi^{2}_{\text{w}}(\psi)}{\partial\psi_{l}}+\frac{\partial\chi^{2}_{\text{s}}(\psi)}
  {\partial\psi_{l}}\stackrel{!}{=}0\;.
\label{chimin}
\end{equation}
The main advantage of the maximum-likelihood approach is its enormous flexibility. In principle, one can incorporate every additional observable constraint that can be connected in some way to the lensing potential. This is of course not restricted to lensing. One simply has to add separate and independent $\chi^{2}$-functions and minimise their sum with respect to the discrete potential values. Even if we are only using weak and strong lensing constraints for now, future improvements of our method should include as many of these constraints as possible.

\subsection{Resolution issues}
\label{resolutionproblems}

\begin{figure}[ht]
\centering
  \includegraphics[width=.89\hsize]{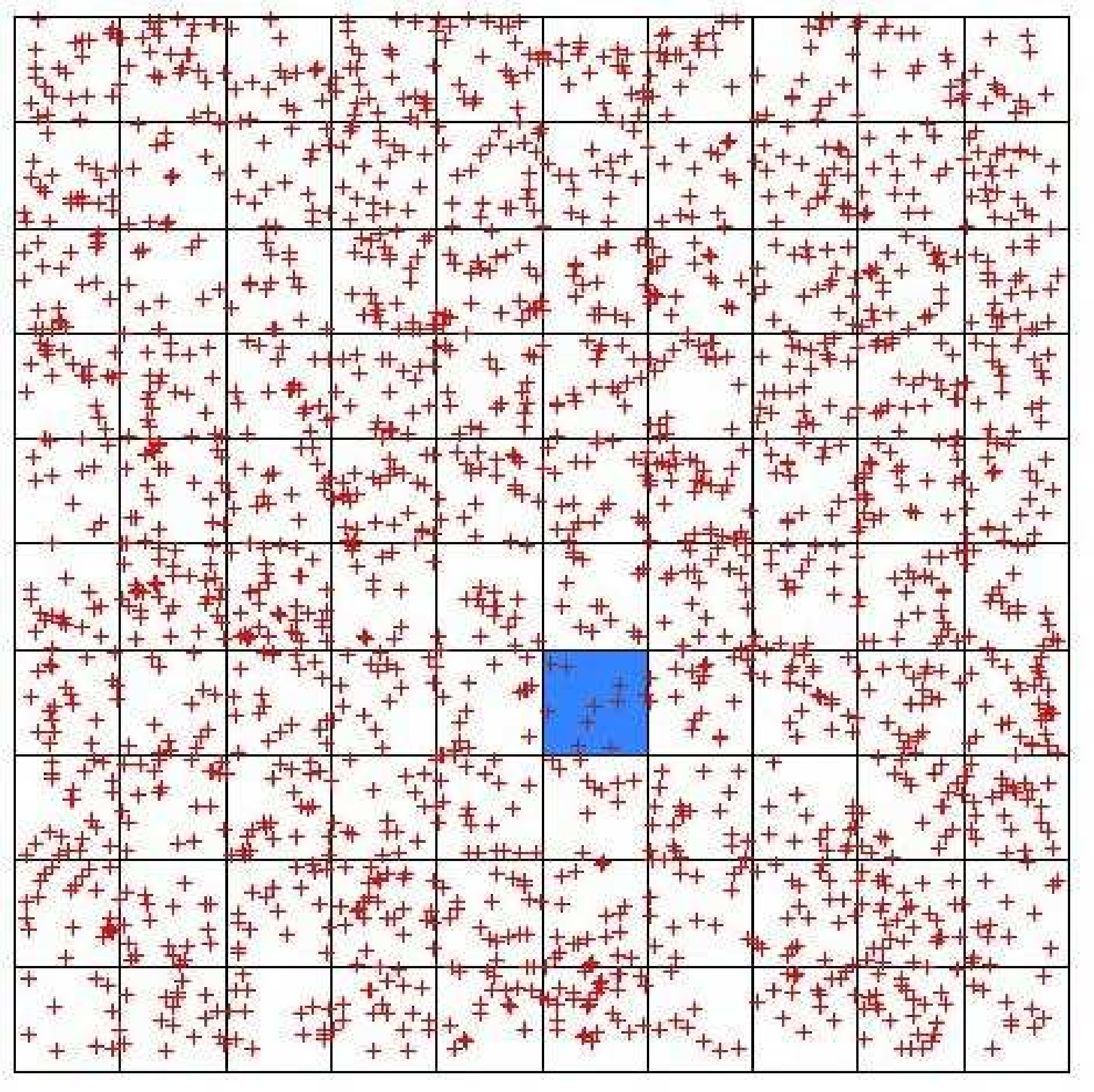}
  \includegraphics[width=.89\hsize]{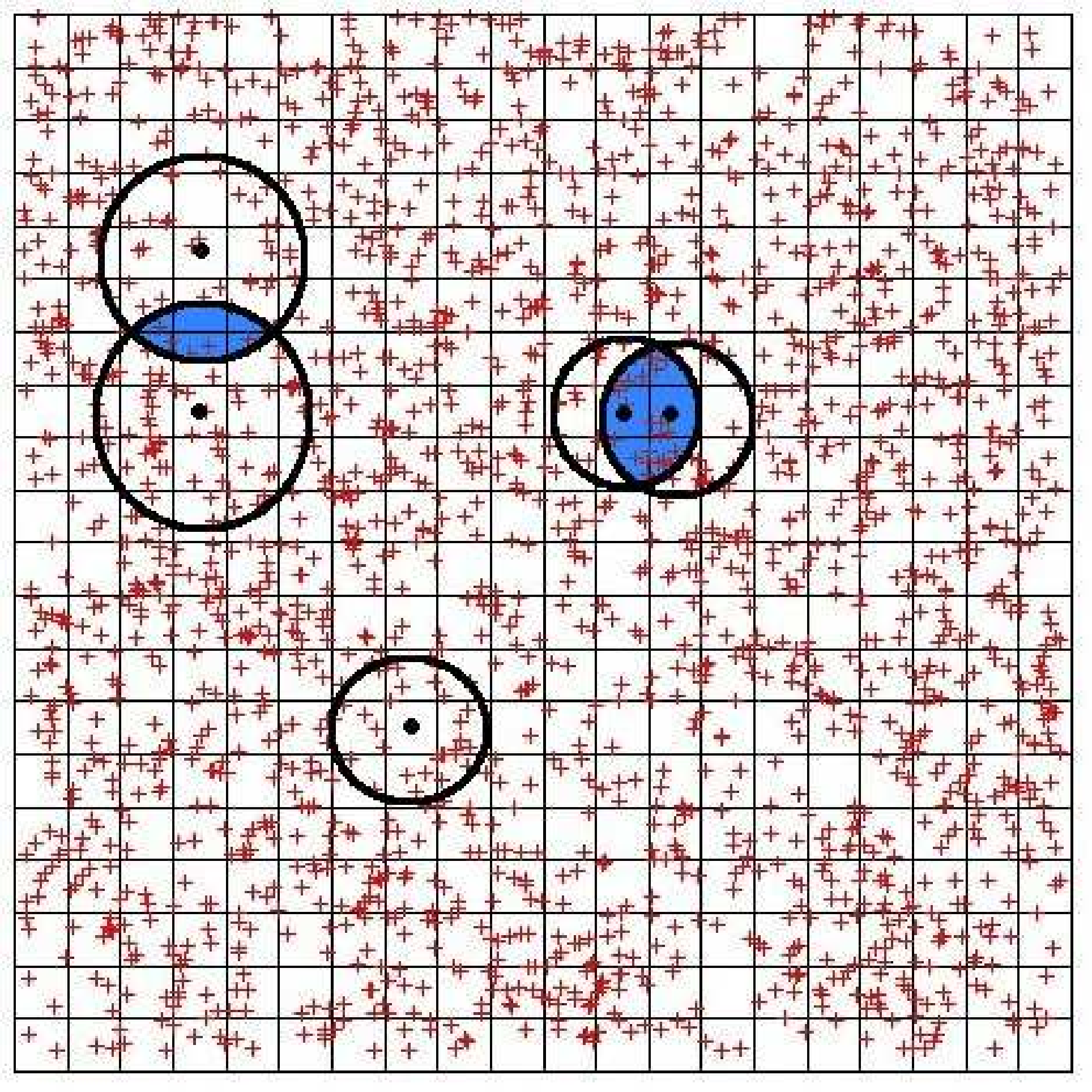}
\caption[Weak-lensing resolutions]{\textit{Top panel}: Very coarse grid of $10\times10$ pixels. An example for a pixel with ten galaxies included is shown in blue. The irregular distribution of galaxies with several void areas is clearly seen. This problem is fixed in the \textit{bottom panel}, showing a grid of $20\times20$ pixels. The circles show the adaptive averaging scales for each individual pixel, which causes overlap, illustrated in blue.}
\label{overlap}
\end{figure}

Before we can start computing the $\chi^{2}$-functions for our joint lensing reconstruction, we have to address two issues concerning the resolution of our reconstruction grid. The first is related to the weak-lensing regime. If we want to average over at least ten galaxies per pixel, the typical background-galaxy density in the field of a cluster would not allow a higher resolution than $\sim10\times10$ pixels, which is of course way too coarse to see any cluster substructures. In addition, pixels of a homogeneous grid can occur which contain fewer than 10 or even no galaxies because of the inhomogeneous, random galaxy distribution.

We solve these problems by an adaptive averaging procedure, in which we average galaxy ellipticities within circles around each pixel centre. Their radii are stepwise increased until each circle contains the desired number of galaxies. Different pixels will need different radii, depending on the local galaxy density in that area of the field. On a fine grid, galaxies shared by neighbouring pixels will of course cause these pixels to be correlated (see Fig.\ref{overlap}).

\begin{figure}[ht]
  \includegraphics[width=0.49\hsize]{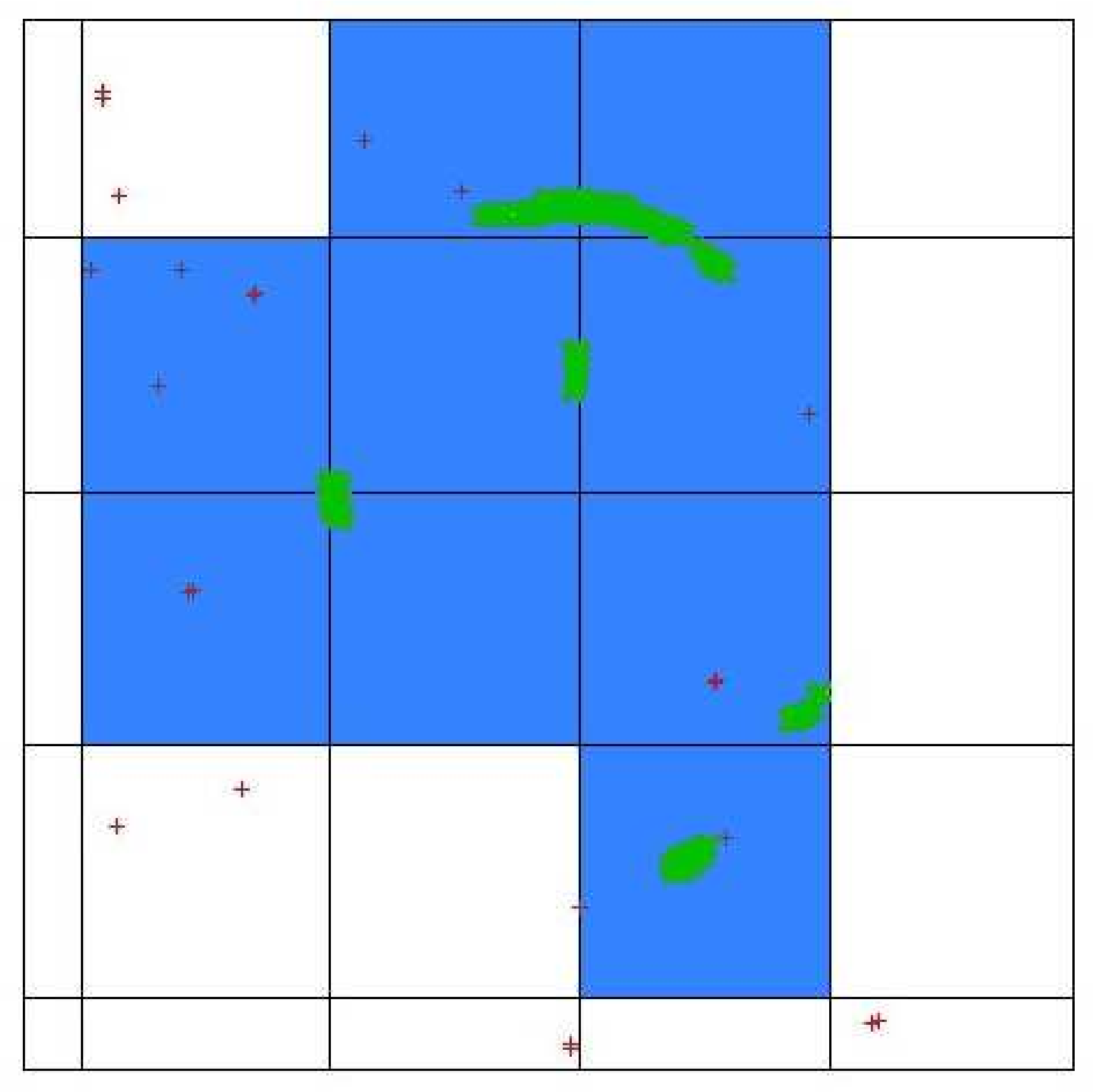}
  \includegraphics[width=0.49\hsize]{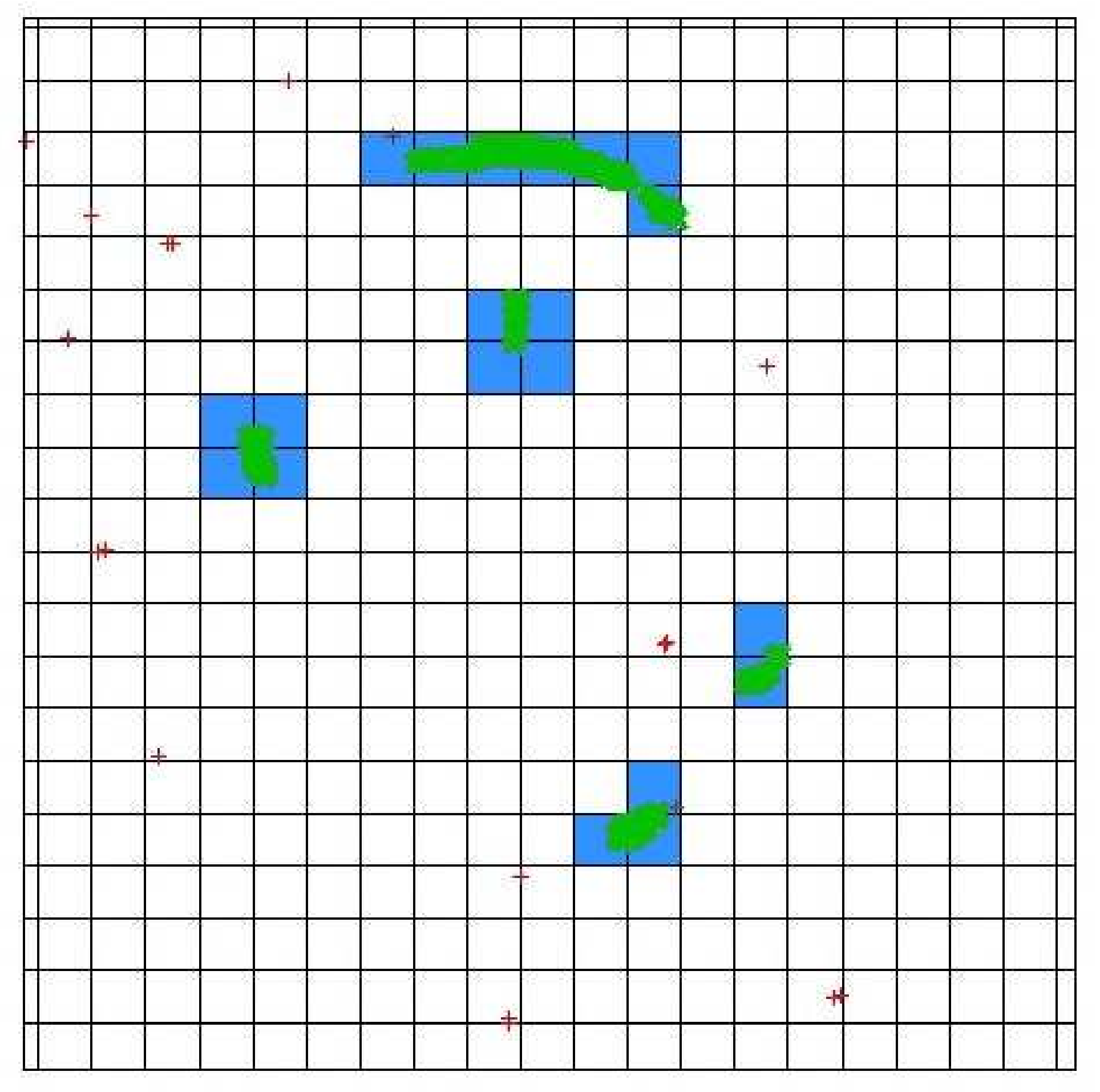}
\caption[Strong-lensing resolutions]{Zoom into the inner part of the cluster. The \textit{left panel} illustrates that a coarse, $20\times20$ grid covering the whole field is by far not able to resolve arc positions. The \textit{right panel} shows a $100\times100$ pixel resolution with respect to the whole field, which is able to follow the arc positions.}
\label{refinedgrid}
\end{figure}

The second issue concerns the strong-lensing regime, in particular the arc positions. Since strong lensing is confined to much smaller scales than weak lensing, essential positional information is lost if the strong-lensing constraints are incorporated at the same resolution as weak lensing. This requires us to refine the grid near cluster centres until it is capable of resolving the exact arc positions (see Fig.~\ref{refinedgrid}).

\subsection{Defining maximum likelihood functions}
\label{definingmaximumlikelihoodfunctions}

The most important ingredient of our cluster reconstruction is the $\chi^{2}$-function (Eq.~\ref{chi2}) that we need to minimise. Consider first the weak-lensing term. As discussed before, the weak-lensing grid pixels are correlated because of the adaptive-averaging procedure, and the expectation value of the ellipticity is the reduced shear rather than the shear. Thus we find, for $|g|\leq 1$,
\begin{equation}
  \chi^{2}_{\text{w}}(\psi)=\left(
    \left\langle\varepsilon\right\rangle -\frac{Z(z)\gamma(\psi)}{1-Z(z)\kappa(\psi)}
  \right)_{i}\mathcal{C}^{-1}_{ij}\left(
    \left\langle\varepsilon\right\rangle -\frac{Z(z)\gamma(\psi)}{1-Z(z)\kappa(\psi)}
  \right)_{j}\;,
\label{wchi}
\end{equation}
where $\left\langle\varepsilon\right\rangle$ represents the results of the averaging process for each pixel. We are using Einstein's sum convention here and below. The case $|g|>1$ is not relevant in our reconstruction because it only affects at most very few pixels on the reconstruction grid.\\
Eq. \ref{wchi} illustrates the first major improvements in our method since \citet{Cacciato2006}. First, we introduce the reduced shear instead of the shear as a reconstruction constraint. Furthermore, we introduce the adaptive averaging technique which adapts to the actual galaxy distribution of the field. Thus, the error in the $\chi^{2}$-function is quantified by the non-diagonal covariance matrix $\mathcal{C}_{ij}$, which we evaluate now. The standard deviation $\sigma_{i}$ for each weak lensing pixel is obtained during the averaging process as the standard deviation from the mean. This standard deviation has three contributions assumed independent,
\begin{equation}
  \sigma=\sigma_{\text{int}}+\sigma_{\text{meas}}+\sigma_{\text{sys}}\;,
\end{equation}
which are the noise due to the intrinsic ellipticity $\sigma_{\text{int}}$, noise introduced by measurement uncertainties $\sigma_{\text{meas}}$ and a systematic noise term $\sigma_{\text{sys}}$ which arises from the fact that the galaxies over which we average cover spatial ranges in which the properties of the lens may change. Here one can also see that the radii of the averaging circles should not become too large, otherwise the coherence in the lensing signal tends to be lost.

Starting from the definition of the covariance matrix,
\begin{equation}
  \mathcal{C}_{ij}=\left\langle
    (x_{i}-\langle x_{i}\rangle)(x_{j}-\langle x_{j}\rangle)
  \right\rangle\;,
\end{equation}
where $x_{i}$ is the ellipticity sample of the pixel with index $i$ and using that the correlation between two pixels due to the averaging process will be proportional to the overlap between the averaging circles attached to them as shown in Fig. \ref{overlap}, we arrive at
\begin{equation}
  \mathcal{C}_{ij}=w_{ij}\sigma_{i}\sigma_{j}\;.
\label{covariance}
\end{equation}
The weight factors $w_{ij}$ are obtained from the number of galaxies contained in the overlap area of both circles,
\begin{equation}
  w_{ij}=\frac{2N_{ij}}{N_{i}+N_{j}}\;,
\label{weighting}
\end{equation}
where $N_{i}$ and $N_{j}$ are the galaxy numbers contained in the circles around pixel centres $i$ and $j$, and $N_{ij}$ is the number of galaxies contained in their overlap. These weights have the expected properties, i.e.~they are unity if $i=j$ and vanish for completely independent and uncorrelated pixels.

The strong-lensing term looks much simpler. By definition, the determinant of the Jacobian vanishes on the critical line. Thus, if we know which pixels are traversed by the critical curve, we can define
\begin{equation}
  \chi^{2}_{\text{s}}(\psi)=\frac{\left(\det{\mathcal A}(\psi)\right)^{2}_{i}}{\sigma^{2}_{\text{s}}}=
  \frac{\left((1-Z(z)\kappa(\psi))^{2}-|Z(z)\gamma(\psi)|^{2}\right)^{2}_{i}}{\sigma^{2}_{\text{s}}}\;,
\end{equation}
where the strong-lensing error estimate $\sigma_{s}$ is mainly caused by the finite pixel size of our grid, since this determines the inaccuracy of the position of the critical curve. We approximate this uncertainty to first order with the help of the Einstein angle \citep[see][]{Cacciato2006},
\begin{equation}
  \sigma_{\text{s}}\approx\left.\frac{\partial\det\mathcal{A}}{\partial\theta}\right|_{\theta_{c}}\delta\theta
  \approx\frac{\delta\theta}{\theta_{E}}\;,
\end{equation}
with the pixel size $\delta\theta$. This expression holds exactly for an isothermal sphere, but can be used as a good approximation for the noise of the critical-curve position.

To evaluate Eq.~(\ref{chimin}), we have to connect convergence and shear to the grid values of the potential, which we shall do in the next section.

\subsection{Lensing potential and convergence maps}
\label{lensingpotentialandconvergencemaps}

The lensing potential which we are reconstructing is not directly observable, but linear combinations of its second derivatives are. Therefore, we can always add a constant, a linear function in $\vec{\theta}$, or a harmonic function to it without changing the observables. Furthermore, if ellipticities are the only quantities measured, the lensing potential is affected by the so-called mass-sheet degeneracy \citep{Falco1985}, which arises because ellipticities are invariant against isotropic scaling of the Jacobian matrix. Note that such transformations also leave the critical curves of a lens unchanged. These degrees of freedom allow the following transformation of the potential \citep{Bradav2004}:
\begin{equation}
  \psi(\vec{\theta},z)\rightarrow
  \psi'(\vec{\theta},z)=\frac{1-\lambda}{2}\vec{\theta}^{2}+\lambda\psi(\vec{\theta},z)\;,
\end{equation}
where $\lambda\ne0$ is an otherwise arbitrary constant.

This is the reason why the reconstructed, discretised potential may look shifted or distorted. However, this is not a problem because we only need its curvature. We obtain  physically meaningful quantities like convergence or shear by simply applying Eqs.~(\ref{convergence}), (\ref{shear1}) and (\ref{shear2}) to $\psi$. We shall use the convergence mainly to describe the reconstruction of a galaxy cluster, because it intuitively reflects the cluster's mass distribution through its surface mass density.

Due to the mass-sheet degeneracy \citep{Falco1985}, the convergence is unique only up to the transformations
\begin{equation}
  \kappa(\vec{\theta},z)\rightarrow\kappa^\prime(\vec{\theta},z)=
  (1-\lambda)+\lambda\kappa(\vec{\theta},z)
\label{massconv}
\end{equation}
with $\lambda\ne0$. If our observed field is sufficiently large, we assume that $\kappa\rightarrow 0$ towards the field boundary, so we can use Eq.~(\ref{massconv}) again for normalisation.

More elaborate methods require observables in the reconstruction which are not invariant under the mass-sheet transformation and depend on potential and convergence. One example is the source magnification, which \citet{Bartelmann1996} suggested to include in the maximum-likelihood approach. Another approach to lift the mass-sheet degeneracy was proposed by \citet{Bradav2004,Bradav2005}, who proposed to exploit the knowledge of the source-redshift distribution.

Having obtained the convergence, possibly transformed according to the mass-sheet degeneracy, mass estimates are straightforward. If we know the lens redshift and fix the cosmological model, we also know the physical area of one pixel. If we additionally know at least the mean redshift of the sources, we can calculate the surface mass density, which yields an estimate for the total cluster mass after summing over the whole grid. Recall, however, that this returns a distance-weighted integral over the entire mass of cosmic structures along the line-of-sight from the observer to the sources.


\section{Implementation}
\label{implementation}

We shall now proceed to the specific implementation and the description of the required numerical methods and algorithms.
As we already pointed out we significantly developed our method with the introduction of an adaptive averaging scheme and the use of the reduced shear instead of the shear. The price that we pay for these improvements is correlated reconstruction pixels and a relatively complicated two-level iteration scheme that we will describe in this section. As a result the runtime of our method increased dramatically which made it necessary to increase the speed of the reconstruction algorithm. The most important step towards speeding up the calculations is the parallisation of our code, using the well-known MPI library 
\subsection{Preparing weak lensing data}
\label{preparingweaklensingdata}

We start with the analysis of the weak lensing data. It is provided in the form of a table containing columns for the position and ellipticity measurement for each distorted background galaxy. It should be noted that the coordinates are arbitrary as long as they all refer to the same coordinate system. We express the coordinates in arcseconds relative to the brightest cluster galaxy (BCG). The reconstruction grid is set up by assigning coordinates to each pixel centre.

The adaptive averaging process proceeds by enlarging circles around each pixel centre until they contain a pre-assigned, constant number of background galaxies. Once this number of galaxies is reached, the average and the standard deviation of the ellipticity are calculated and assigned to the pixel.

The covariance matrix between two pixels is determined by the number of galaxies shared between them. Its final entries are obtained using Eq.~(\ref{covariance}), because we can now calculate the weightings $w_{ij}$ (see Eq.~\ref{weighting}). This procedure has to be done for both ellipticity components, and the resulting covariance matrices must be inverted.

\subsection{Preparing strong lensing data}
\label{preparingstronglensingdata}

\begin{figure}[ht]
  \includegraphics[width=0.49\hsize]{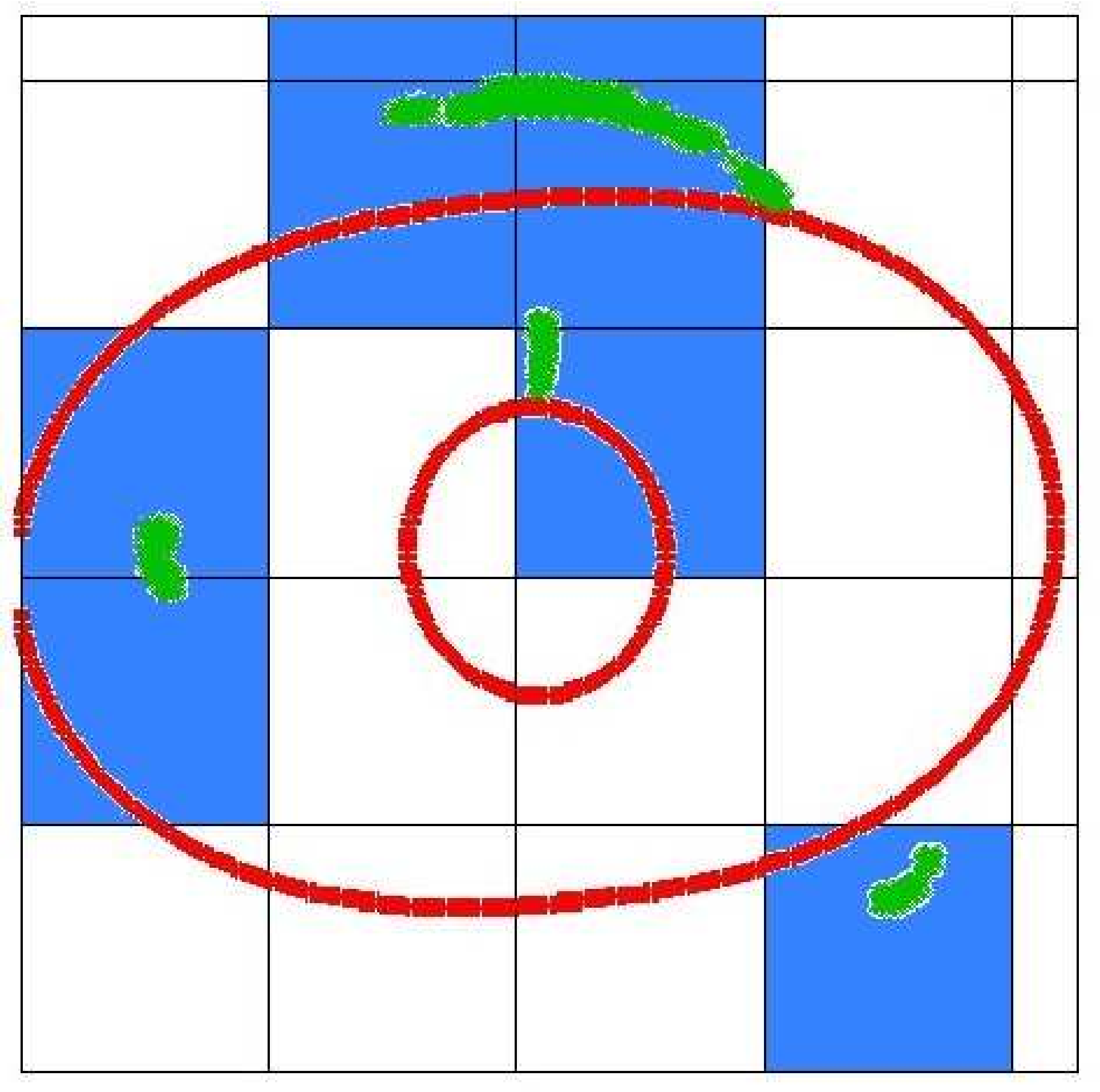}
  \includegraphics[width=0.49\hsize]{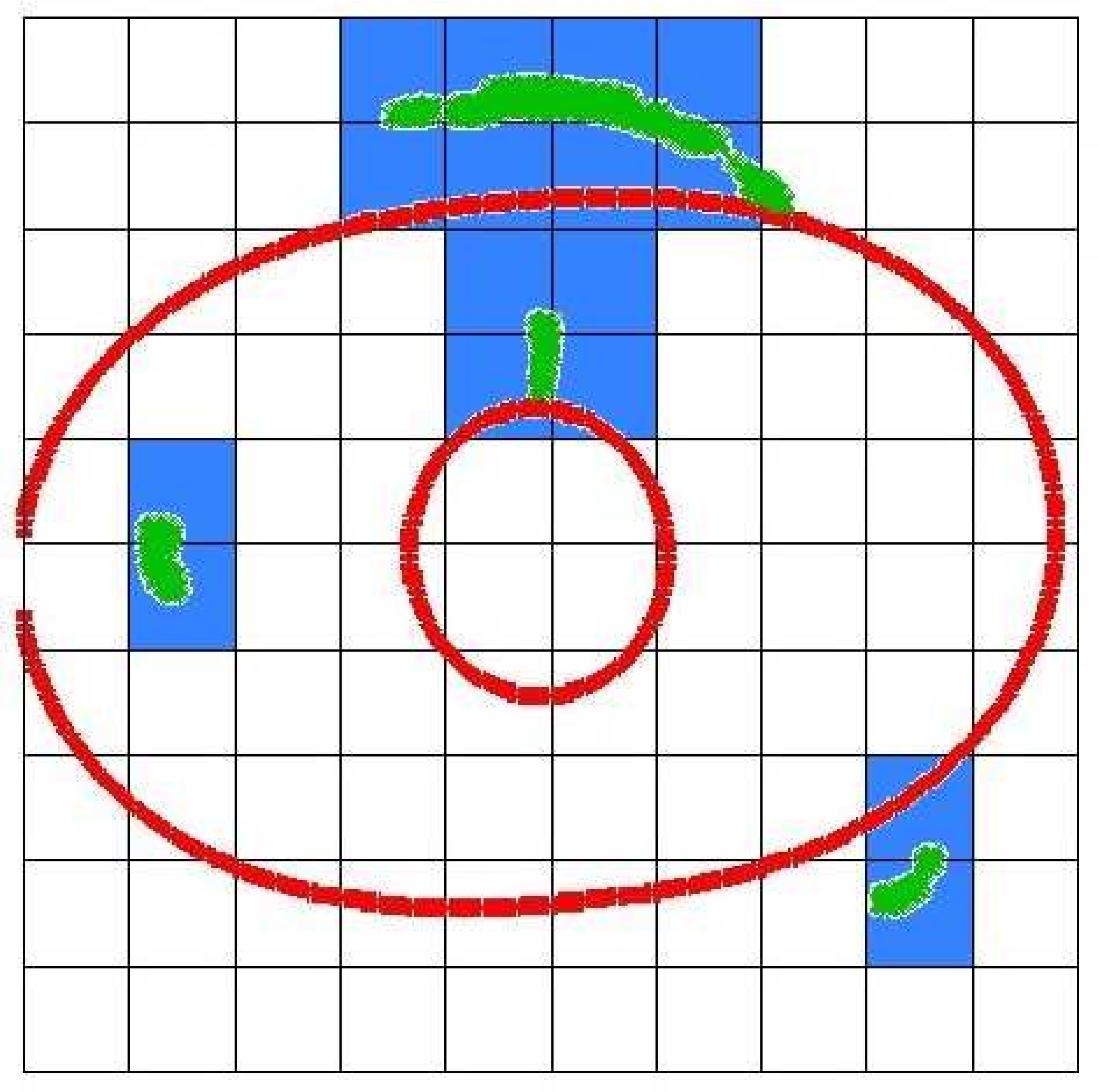}
\caption{Arc position and estimated position of the critical curves for the real cluster MS 2137. The critical curves were estimated by a parametric strong lensing reconstruction from \citet{Comerford2006}. \textit{Left panel} shows 32x32 grid resolution where one can see no deviation. \textit{Right panel} shows 75x75 where one sees that the deviation in position is still small.}
\label{critarc}
\end{figure}

Handling the strong-lensing data, we have to cope with the fact that we cannot observe the critical curves directly. We thus need a good approximation for their locations, which is given by arc positions that can be observed very well. We show in Fig.~\ref{critarc} that arcs follow the position of the critical curves, as long as the resolution of the grid is not extremely high. However, even at  the higher resolution of the finely resolved central grid, the difference in the pixel positions between arcs and critical curves is at most two pixels. \citet{Cacciato2006} showed that deviations of this size do not affect the reconstruction significantly.

A more severe problem is that the arcs sample the critical curves only very sparsely. We cannot expect to obtain full knowledge of the critical curve through observations. It will be one aim of future work to use high-resolution observations of cluster fields which tend to show more strongly lensed images and thus allow tracing of the critical curve in more detail. Another possibility would be to rely on critical-curve reconstructions from parametric strong-lensing analyses, and to feed that critical curve into the code. The drawback of this approach is that one gives up the completely non-parametric nature of the reconstruction by using profile assumptions in the critical-curve determination. In either case, the critical curves are characterised by a table listing the approximate positions of critical points.\\
We finally point out that we are using arcs as approximate indicators for critical-curve locations rather than multiple-image systems as strong-lensing constraints. This is for several reasons, first of all the identification of multiple image systems is not always possible since it requires multi-color observations and the subtraction of cluster members to look for hidden images. Moreover, due to resolution issues the reconstruction cannot be extremely accurate in the cluster centre and as a result we do not expect large changes by using multiple images instead of critical points. \\
But nevertheless one should use as many constraints as possible, which is the reason why future versions of our method will also contain multiple-image system information, if available, to give an optimal reconstruction result.	

\subsection{Grid methods}
\label{gridmethods}

We now proceed to combine lensing theory and the measurements, and to implement the method numerically.

We defined the $\chi^{2}$-functions in Sect.~\ref{definingmaximumlikelihoodfunctions} using the discretised lensing-potential values as minimisation parameters. The observable to be reproduced in the case of weak lensing is the reduced shear, and the location of a vanishing determinant of the Jacobian in the case of strong lensing.

In order to minimise the resulting $\chi^{2}$-function with respect to the lensing potential, we have to write the convergence and the shear in terms of the discrete potential values. The relation between these quantities is given by Eqs.~(\ref{convergence}), (\ref{shear1}) and (\ref{shear2}) with the second derivatives replaced by finite-differencing schemes.

\begin{figure*}[ht]
  \includegraphics[width=\hsize]{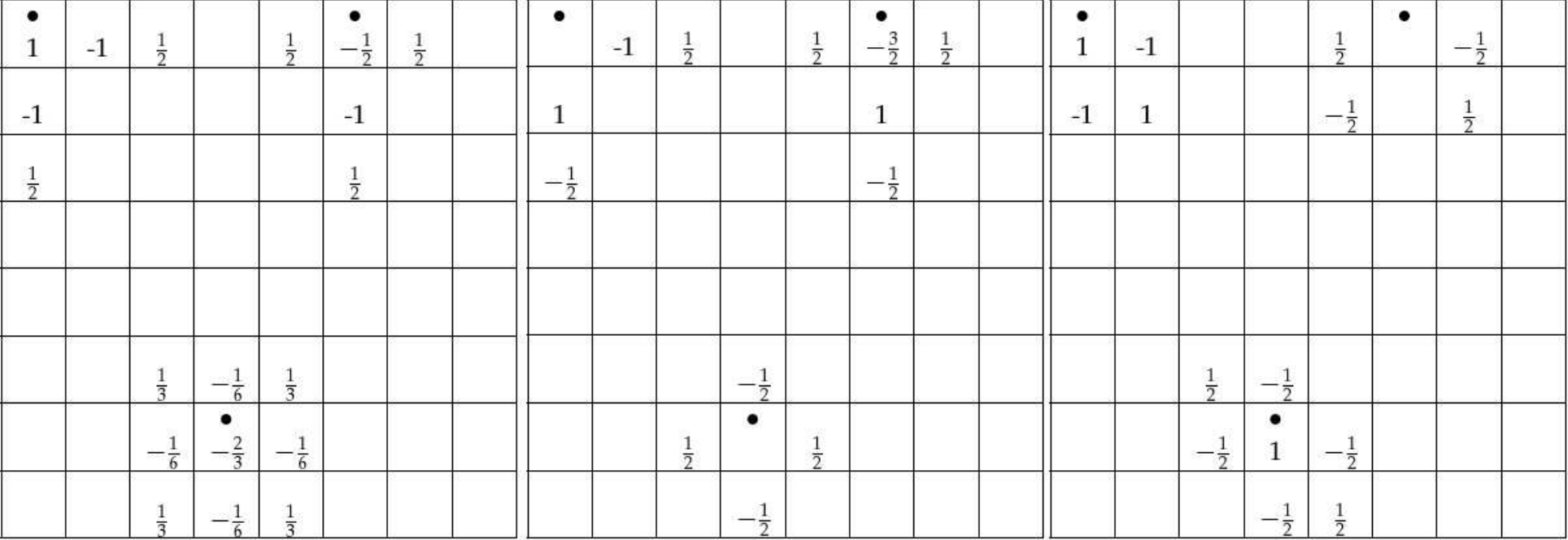}
\caption{Finite differences schemes. Written in the cells are the coefficients in the difference quotients. \textit{Left panel}: Scheme for the convergence; \textit{centre panel}: scheme for the first shear component; \textit{right panel}: scheme for the second shear component.}
\label{findifschemes}
\end{figure*}

Given a certain grid resolution, we want to obtain the lensing potential at each grid position $(x,y)$ with $1\le x\le M$, $1\le y\le N$. We enumerate these grid positions sequentially line by line and use the central difference quotients for discrete representations of the second derivatives at a given grid position. Our finite differencing scheme is identical to that chosen by \citet{Cacciato2006}, and the corresponding coefficients are given in Fig.~\ref{findifschemes}. At the edges and borders of the grid different (one-sided) finite-differencing schemes need to be used.

With our way of enumerating the pixels, we can write the discrete potential $\vec{\psi}$ on the $N\times M$ grid in the form of a vector $\vec{\psi}\in \mathbb{R}^{NM}$. This only slightly complicates addressing the correct positions in the vector. Direct left and right neighbours are just separated by $\pm1$ positions in the vector, while top and bottom neighbours are separated by $\pm N$ positions. The advantage of this enumeration is that the finite differencing becomes a simple matrix multiplication,
\begin{eqnarray}
\label{Kij}
  \kappa_{i}&=&\mathcal{K}_{ij}\psi_{j} \\
\label{G1ij}
  \gamma^{1}_{i}&=&\mathcal{G}^{1}_{ij}\psi_{j} \\
\label{G2ij}
  \gamma^{2}_{i}&=&\mathcal{G}^{2}_{ij}\psi_{j}\;.
\end{eqnarray}
Here, $\mathcal{K}_{ij}, \enspace \mathcal{G}^{1}_{ij}, \enspace \mathcal{G}^{2}_{ij}$ are sparse band matrices encoding the information on the finite-differencing scheme. The fact that we know these matrices perfectly is the key point to increase the speed of our reconstruction algorithms, which plays an important role with respect to runtime, besides the parallelisation of the code. Without several numerical tricks the runtime of a complete reconstruction would not stay at an acceptable level.

Using these finite-differencing schemes, $\chi^{2}$ can be minimised by solving the linear system of equations
\begin{equation}
  \mathcal{B}_{lk}\psi_{k}=\mathcal{V}_{l}\;,
\end{equation}
where the coefficient matrix $\mathcal{B}_{lk}$ and the result vector $\mathcal{V}_{l}$ contain on the one hand observed ellipticity and critical curve information and on the other hand the convergence and shear matrices. The calculation is detailed in the Appendix.

\subsection{Regularisation and 2-step-iteration}
\label{regularisationand2stepiteration}

\subsubsection{Regularisation}
\label{regularisation}

We now introduce a regularisation term $R(\psi)$ into the $\chi^2$ function  Eq.~(\ref{chi2}) to obtain
\begin{equation}
  \chi^{2}(\psi)=\chi^{2}_{\text{w}}(\psi)+\chi^{2}_{\text{s}}(\psi)+R(\psi)\;.
\label{R}
\end{equation}
The regularisation term depends on the potential and is defined such as to disfavour unwanted solutions. This is necessary to prevent the reconstruction from following intrinsic noise patterns, which do not reflect intrinsic features of the true underlying potential. Here, we use a simple regularisation term scheme simply comparing the convergence obtained with that found in the preceding iteration,
\begin{equation}
  R=\eta_{i}\left(\kappa_{i}^{\text{before}}-\kappa_{i}(\psi)\right)^{2}\;.
\label{eta}
\end{equation}
Its amplitude $\eta_{i}$ controls the relative strength of the regularisation and is crucial for the reconstruction. It should be chosen such that the overall $\chi^{2}$ is of order unity and of course low enough for changes in the reconstruction to take effect. Minimising the regularised $\chi^2$ of Eq.~(\ref{eta}) again leads to a linear system which contributes one additional term to the coefficient matrix and the result vector. The complete calculation is presented in the Appendix. In our reconstruction we also regularize on the two shear components which give similar terms. Note that it is not useful to directly regularize the potential since it looks very different in each iteration step, as described in Sect.~\ref{lensingpotentialandconvergencemaps}.

\subsubsection{Inner-level iteration}
\label{innerleveliteration}

As the full expressions for the $\chi^{2}$ minimisation show, it is not precisely a linear system, but includes non-linear prefactor terms in the coefficient matrix and the result vector. We solve this problem in the same way as \citet{Bradav2005} did by means of an iterative approach in which the non-linear terms are computed from the preceding iteration. Starting from a first guess for the convergence, we express the corresponding non-linear terms by constant factors, as can be seen in the Appendix.

\citet{Bradav2005} showed that the initial guess of the convergence does not affect the final reconstruction, but at most the number of iterations. We confirm their result, which implies that the initial guess of a flat convergence is appropriate. We minimise $\chi^2$ and obtain a solution for the lensing potential. From it, we calculate convergence and shear, which we insert as new guesses into the non-linear terms in the next step of the reconstruction. This yields a convergent iterative process. We control the convergence of the procedure by comparing its results between subsequent iterations. If the change falls below a given threshold, we stop the iteration. The drawback of this method is that we now need several iterations (3-5 in practice) for a complete reconstruction, which takes some time at high resolution.

\subsubsection{Outer-level iteration}
\label{outerleveliteration}

The regularisation term also helps in a second type of iteration, which we call outer-level iteration. The background-galaxy density of today's observations allows a resolution of only $\sim10\times10$ uncorrelated pixels. Higher resolution is desirable at the expense of correlated pixels, which will affect the reconstruction. Also the convergence values in some individual pixels will increase at higher resolution, which renders the initial guess of a vanishing convergence increasingly less accurate, giving rise to many inner-level iterations.

These problems can easily be avoided by introducing another iteration as follows:

\begin{itemize}

\item One begins at the lowest possible resolution, where the pixels are not or almost not correlated. This resolution will be too coarse for fine-structure noise patterns to appear.

\item Starting from the initial convergence set to zero, the inner-level iteration is performed until the reconstruction converges.

\item The obtained result for the potential is then interpolated by a bicubic spline algorithm to a slightly higher resolution.

\item This interpolation is taken as the comparison function in the regularisation and as a new initial guess in the inner-level iteration at the higher resolution.

\item This process is repeated until the final resolution is reached.

\end{itemize}

\begin{figure}[ht]
  \includegraphics[width=0.49\hsize]{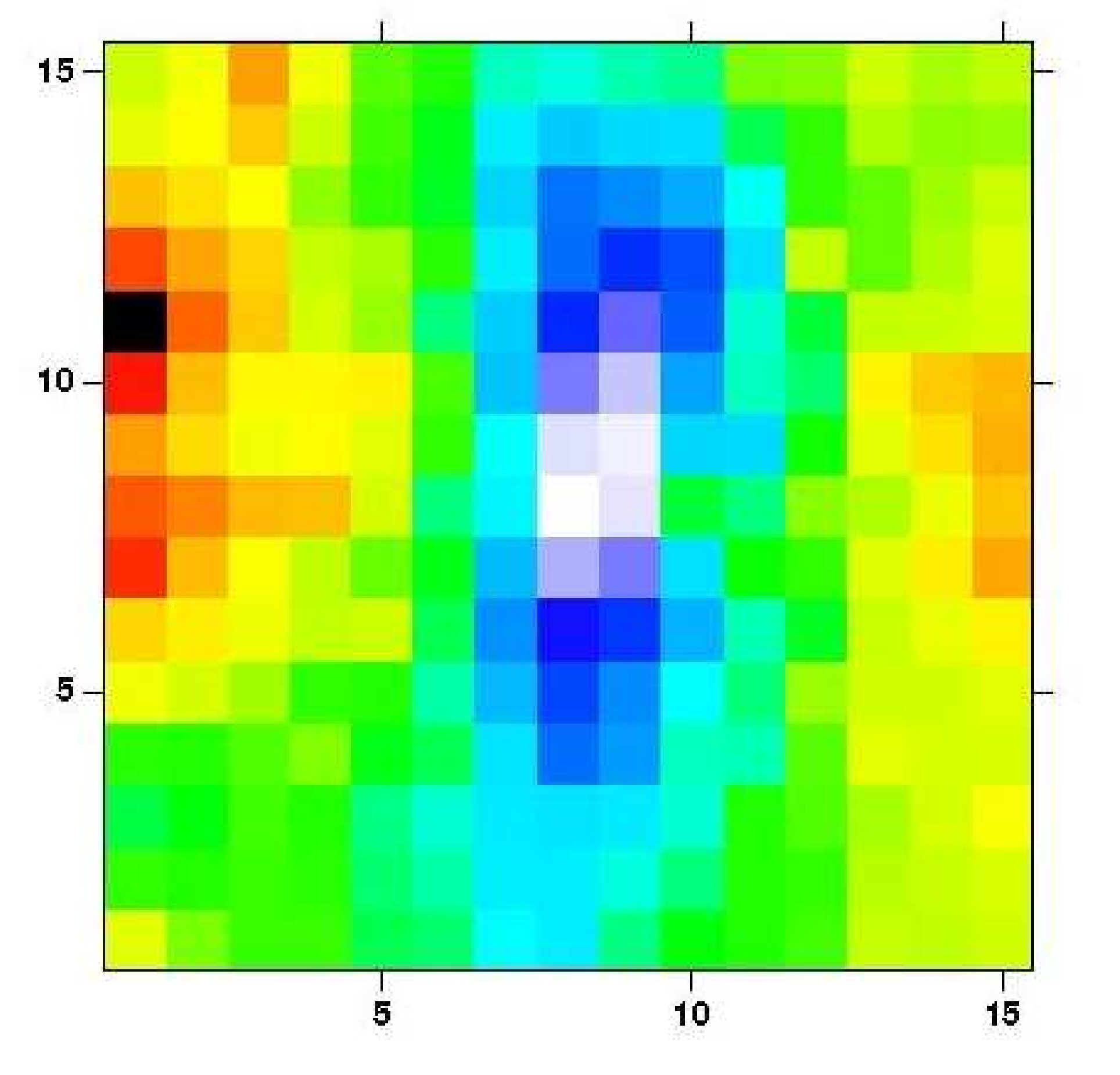}
  \includegraphics[width=0.49\hsize]{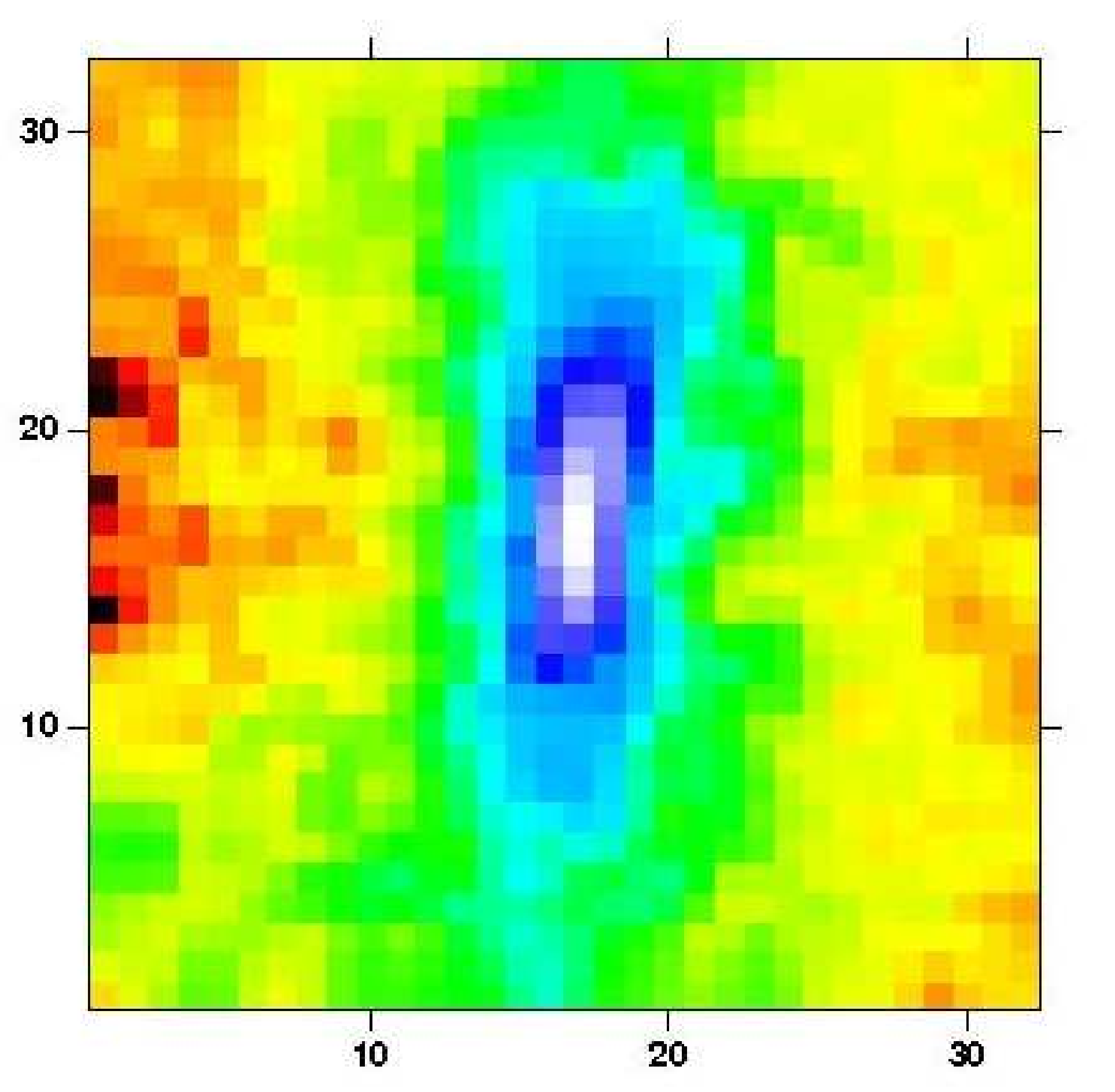}
\caption{Convergence maps of a simulated cluster. \textit{Left panel}: Initial reconstruction starting at a resolution of $15\times15$ pixels. \textit{Right panel}: Final resolution of $32\times32$ pixels, reached after several iterations.}
\label{outergraphic}
\end{figure}

This two-level iteration delivers by far the best reconstruction results, but with the disadvantage of increased CPU time due to the higher number of iterations. The result of the gradual transition from low to high resolution can be seen in Fig.~\ref{outergraphic}.

\subsection{Strong lensing at high resolution}
\label{stronglensingonhighresolution}

\subsubsection{Interpolation}
\label{interpolation}

Even when a weak-lensing reconstruction as described in the previous sections has converged, the grid resolution is far from being fine enough to follow the shape of the critical curves. We deal with the problem by interpolating the lensing potential from the final weak-lensing reconstruction and calculate convergence and shear at that refined resolution. Then, we zoom into the cluster core where the strong-lensing information is available. The interpolation is done by a bicubic spline interpolation routine which is reliable enough to create just a small amount of additional noise. The interpolated result serves as template for the following high-resolution reconstruction.

\begin{figure}[ht]
\centering
  \includegraphics[width=0.89\hsize]{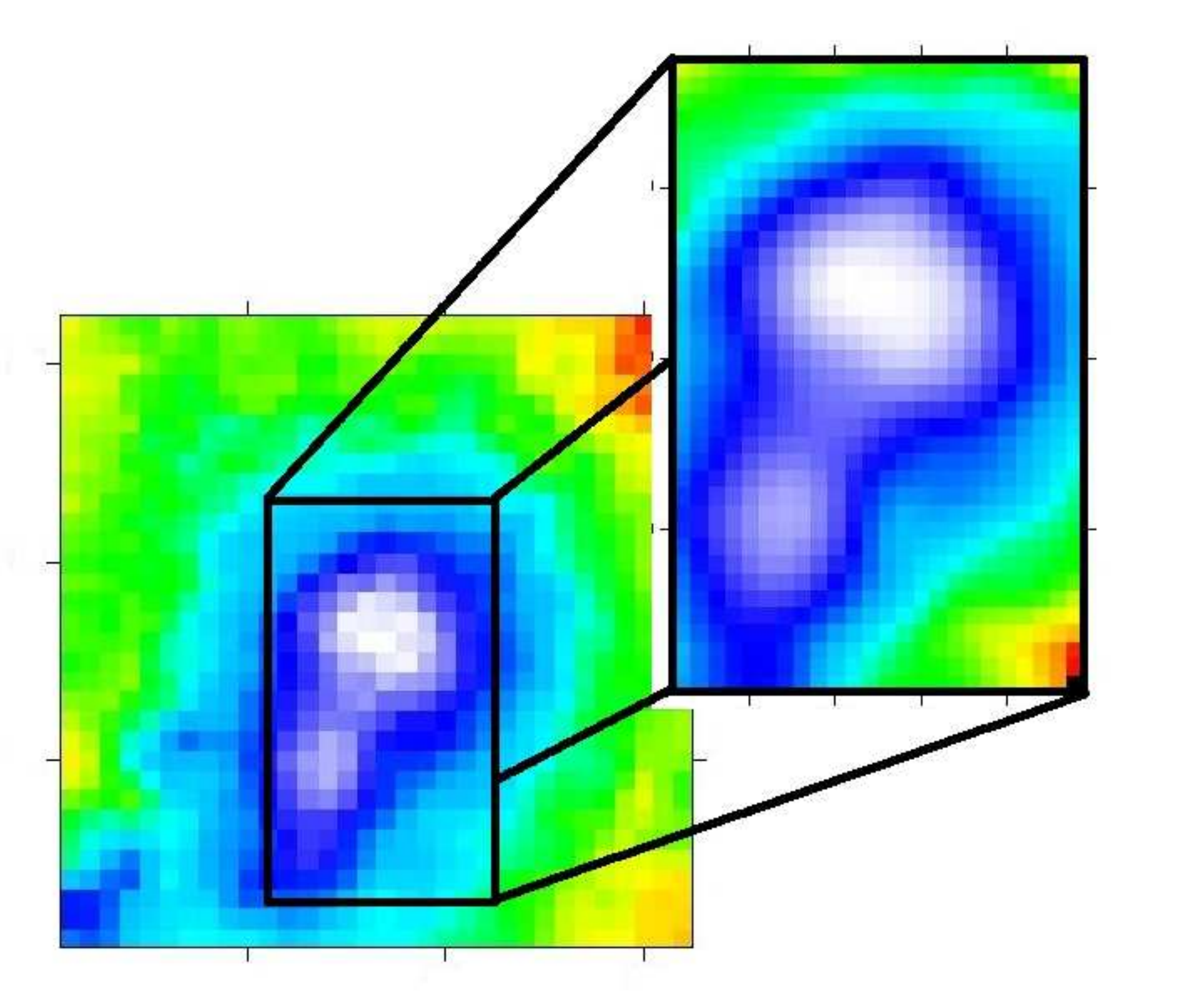}
\caption{\textit{Left panel}: Complete combined weak and strong lensing reconstruction at low resolution of a simulated cluster. \textit{Right panel}: Zoom into the interpolated cluster core of the low-resolution reconstruction.}
\label{interpolzoom}
\end{figure}

\subsubsection{Modified maximum likelihood function}
\label{modifiedmaximumlikelihoodfunction}

Now, we have to remove the weak-lensing term from the $\chi^2$ function,
\begin{equation}
  \chi^{2}(\psi)=\chi^{2}_{\text{s}}(\psi)+R(\psi)\;,
\end{equation}
because no weak-lensing data points are available at the high resolution required in the core. The strong-lensing data points, however, are available at high resolution, because the critical points can be determined very accurately. Yet, the weak-lensing reconstruction enters even at this resolution level through the regularisation term. Thus, the result is based on the weak-lensing constraints.

We finish the reconstruction by inserting the  high-resolution cluster-core result into the result obtained at a coarser resolution, which consists of the complete cluster field. Due to the regularisation function the strong-lensing result fits nicely into the weak-lensing results since we do not allow that the two different reconstructions differ significantly at pixels where no strong-lensing constraints are available. This is also the last major change in our method compared to \citet{Cacciato2006}. The use of the regularisation function as a tool to match results on different scales improves the quality of our reconstructions significantly.


\section{Results}
\label{results}

\subsection{Tests with synthetic data}
\label{synthetictests}

\begin{figure}[ht]
  \includegraphics[width=0.49\hsize]{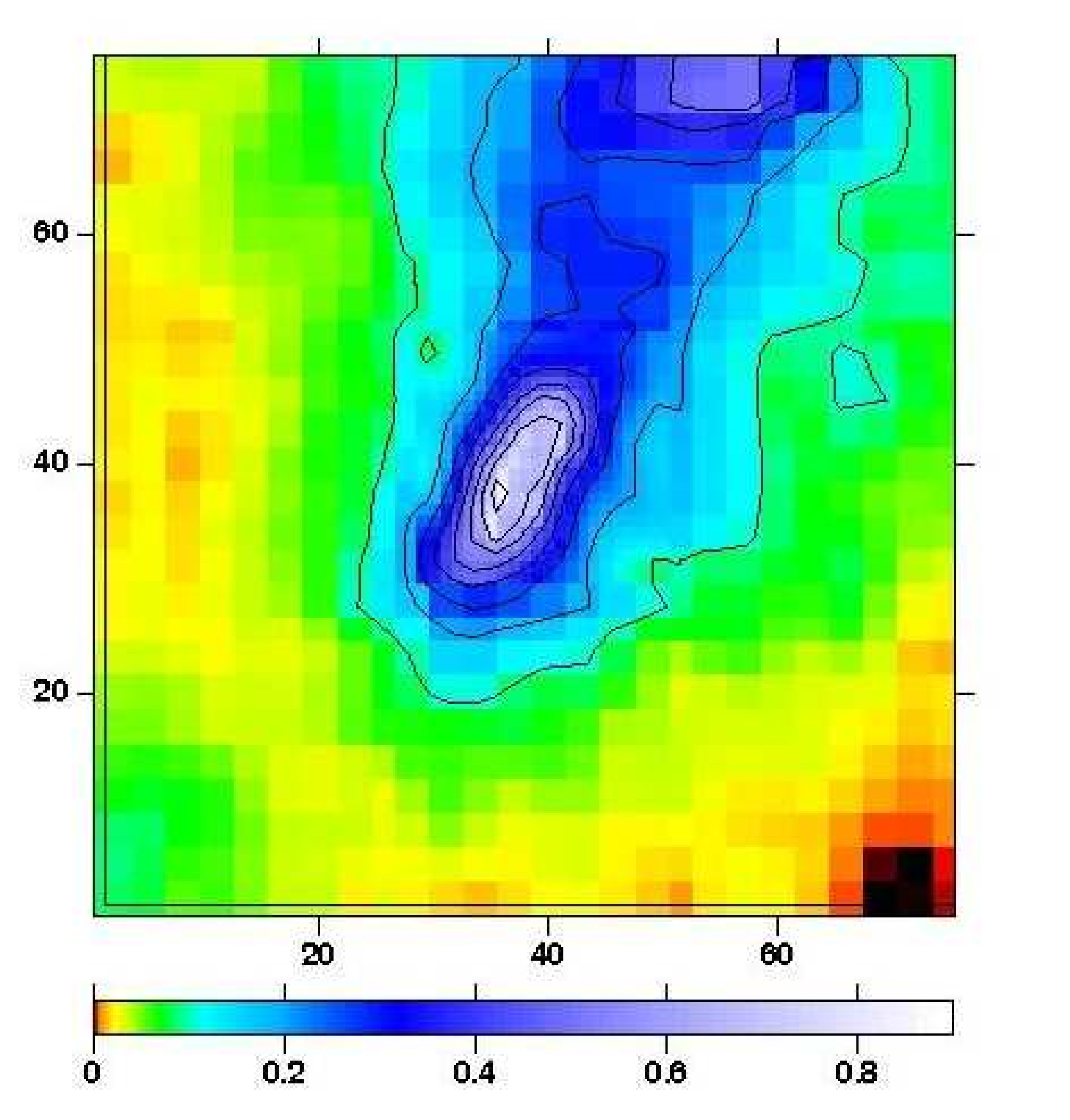}
  \includegraphics[width=0.49\hsize]{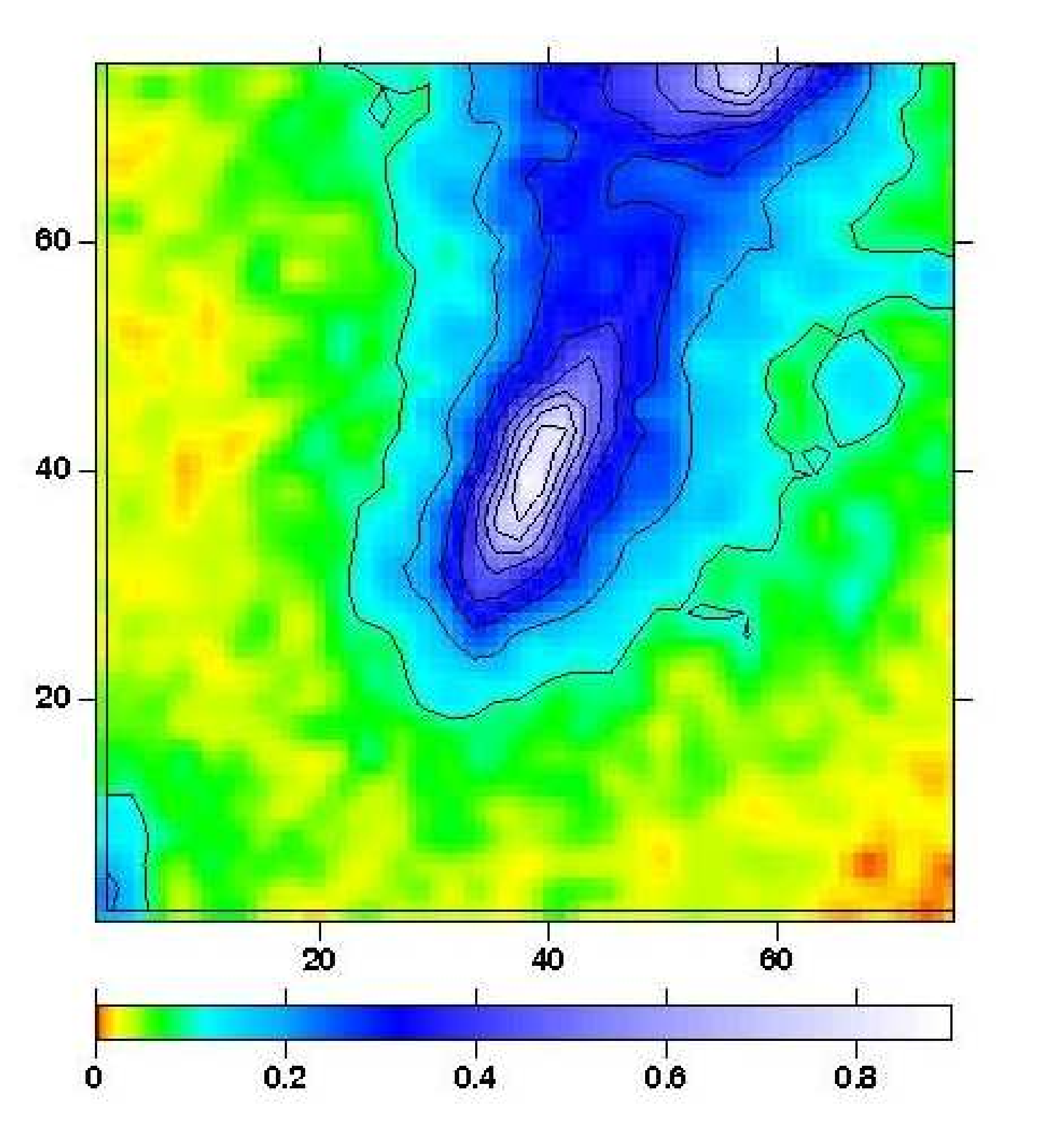}
\caption{\textit{Left panel}: Convergence map of the high-resolution reconstruction. As one can see, the resolution is much higher in the inner part of the map. The colour scale is logarithmic and the contours start at $\kappa=0.1$ spaced by $\Delta\kappa= 0.08$. \textit{Right panel}: The convergence map of the original cluster rebinned at the resolution of the reconstruction. The colour scale and the contour levels are identical in both panels.}
\label{syntheticrec}
\end{figure}

\begin{figure}[ht]
  \includegraphics[width=\hsize]{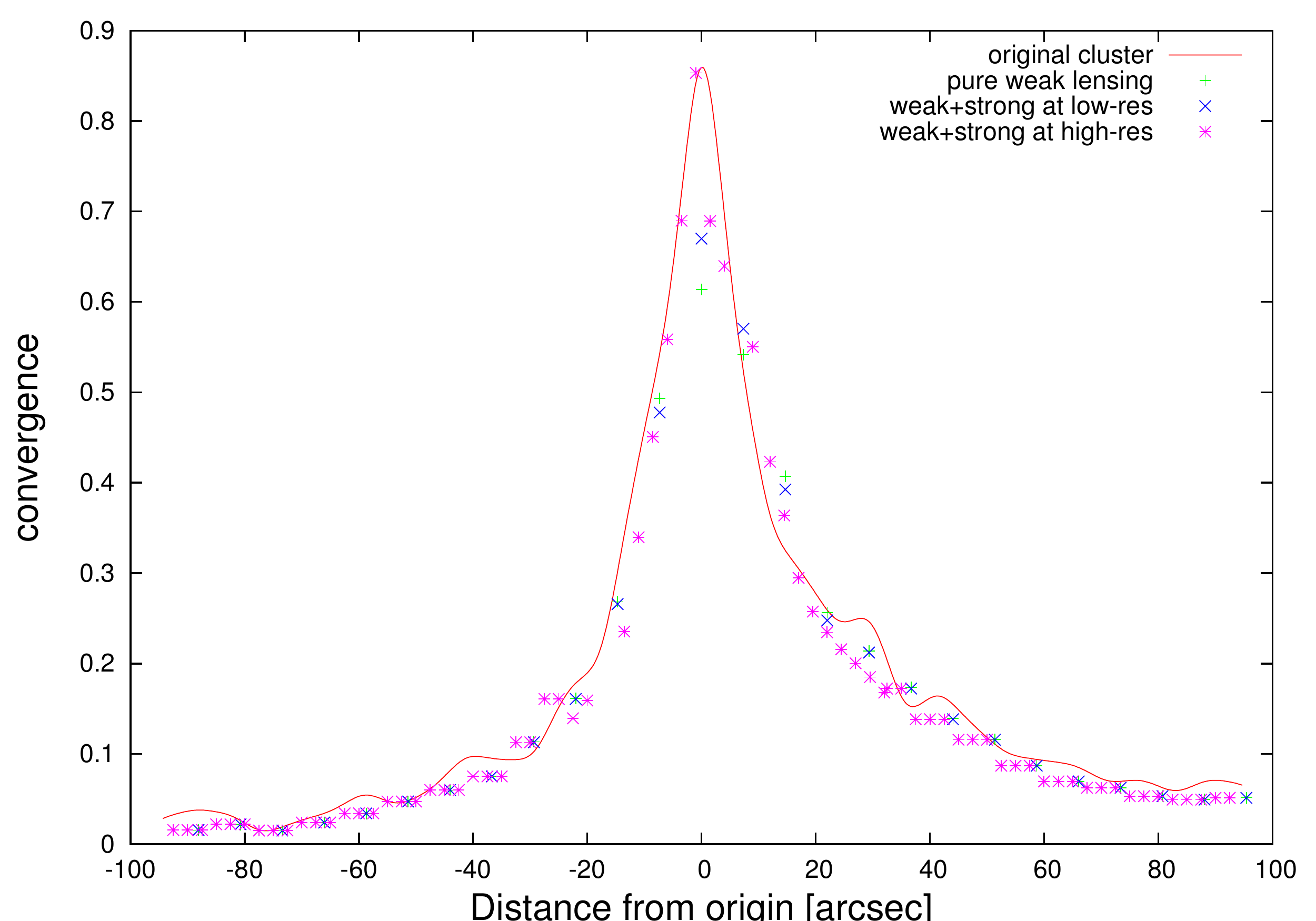}
\caption{The radial $\kappa$ profile along the main diagonal of the critical cluster field.}
\label{profile}
\end{figure}

We first repeat the tests also carried out in \citet{Cacciato2006}. We take simulated clusters from the $N$-body simulations described in \citet{Bartelmann1998} and compute maps of their reduced shear and their critical curves. Based on this information, we try to reconstruct the known potential of the simulated cluster. Since this is an idealised lensing scenario which does not include a realistic background-galaxy distribution or image analysis, it is sufficient to compare the convergence map obtained by the reconstruction with the real convergence map of the simulated cluster. The results confirm the reliability of our method and are shown in Figs.~\ref{syntheticrec} and \ref{profile}. In particular, Fig.~\ref{profile} shows how significantly the results are improved when the strong-lensing constraints on a refined grid are added. Otherwise the central density peak is underestimated by almost 20\%.

\subsection{The lensing simulator}
\label{thelensingsimulator}

\begin{figure*}[ht]
  \includegraphics[width=0.49\hsize]{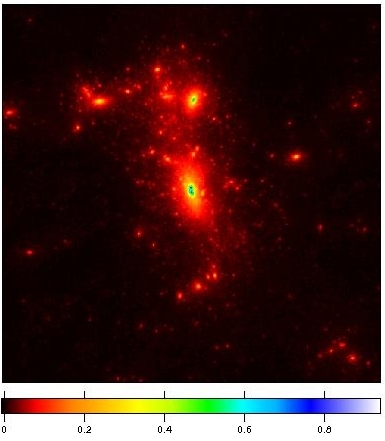}
  \includegraphics[width=0.49\hsize]{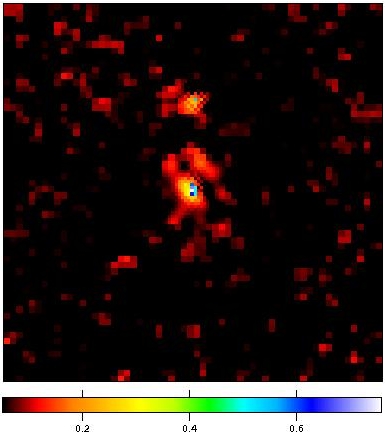}
\caption{\textit{Left panel}: Real convergence map of the simulated cluster.  \textit{Right panel}: Convergence map of the high-resolution reconstruction. The colour scale in both maps is linear and the sidelength corresponds to $\sim10 ~\text{Mpc}$.}
\label{simulation_rec}
\end{figure*}

Next, we use the method detailed in \citet{Meneghetti2008} to simulate an observation of another cluster field.
The target of this simulated observation is the galaxy cluster g72, described in several previous papers \citep{Dolag2005,Puchwein2005,Meneghetti2007a}. This cluster was simulated with different physics. For the present work we have used a pure dark matter simulation. The cluster is at redshift $z_{\text{c}}=0.297$ and has a main halo mass of ~$M_{200}=6.7\times10^{14}M_{\odot}/h$.~It is in the process of merging with a massive substructure of mass~$M_{200}\sim3\times10^{14}M_{\odot}/h$. In the projection chosen for this simulation the subclump is located at $\sim 150~\text{kpc}/h$~north of the main clump. A second massive substructure is present at a distance of $\sim2.5~\text{Mpc}/h$~from the cluster centre. The projected density of the cluster is shown in Fig. \ref{simulation_rec}.\\
We mimic a $2500^{\prime\prime}\times2500^{\prime\prime}$~SUBARU observation of this cluster and of the sky behind it in the R-band assuming an exposure time of 6000s and an isotropic, Gaussian PSF of~$0.6^{\prime\prime}$~FWHM. The distortion field, which is used to lens the background galaxies, is calculated from the cluster mass distribution following the method described in \citet{Meneghetti2007a}. The background galaxies have realistic morphologies, being drawn from shapelet decompositions of real galaxies taken from the Hubble-Ultra-Deep Field (HUDF). Their luminosity and redshift distributions also reflect those of the HUDF \citep{Coe2006}.\\
The weak lensing analysis of the field was carried out by F. Bellagamba (Univ. of Bologna) using an advanced KSB method \citep{Kaiser1995}. It returned an ellipticity catalogue of 39788 background sources.
For the strong lensing analysis we followed two different approaches. First, we used the known complete critical curves to obtain a result under optimal conditions. In a second reconstruction we choose a kind of worst-case scenario where we used four estimated points of the critical curve. These point were determined following an observational approach. First, we identified in the simulation a few fold arcs. Then, we isolated the brightest knots along the lensed images, which are used to constrain the position of the critical lines passing in between them.

\begin{figure}[ht]
  \includegraphics[width=\hsize]{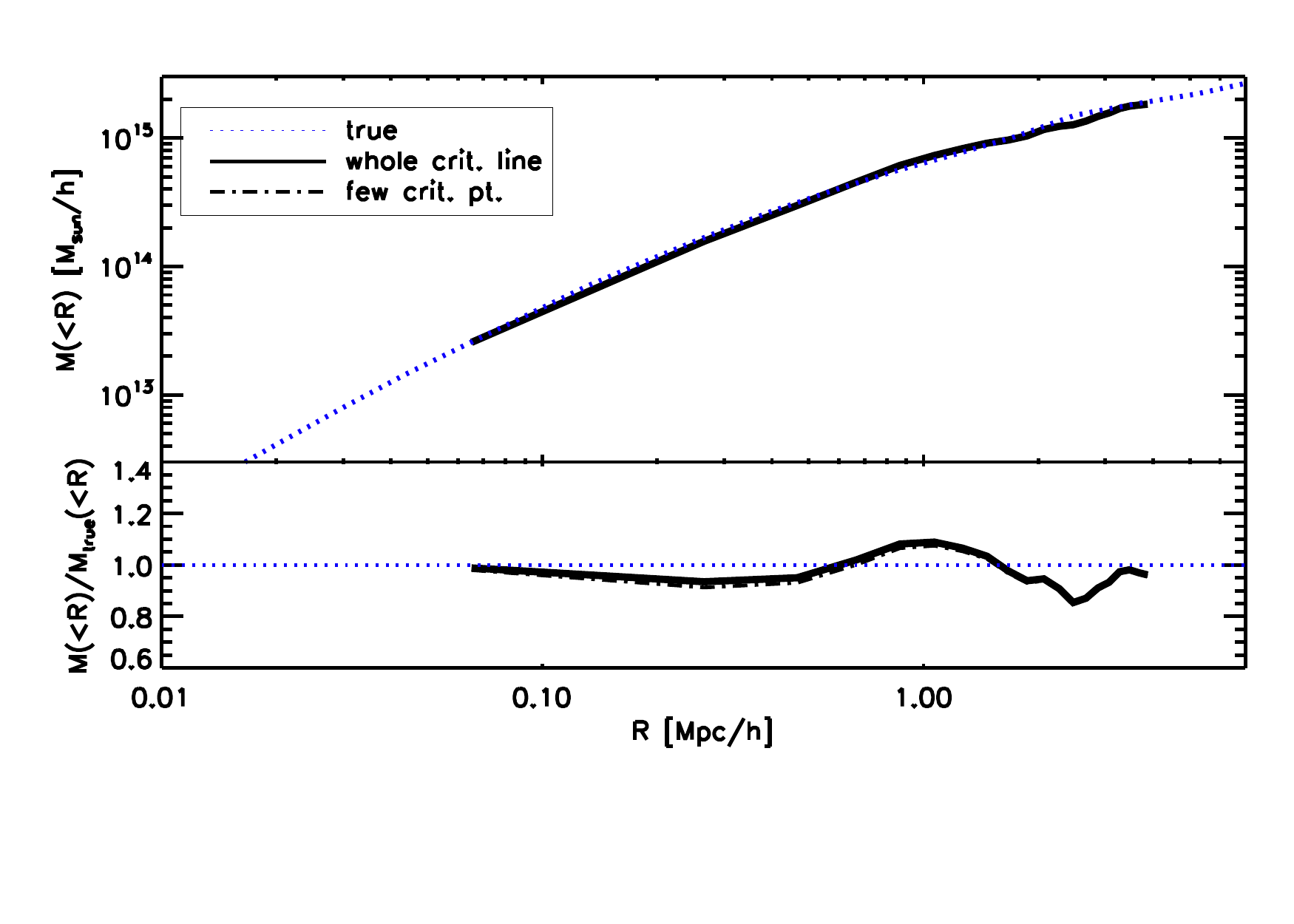}
\caption{\textit{Top panel}: The real and the reconstructed total mass of the simulated cluster within a certain radius. \textit{Bottom panel}: The residuals from the real mass at different scales. Both panels show both reconstructions, based on the full critical-curve knowledge and on only a few points of the critical curve respectively.}
\label{simulation_comp}
\end{figure}

A first qualitative look at the obtained convergence map  already shows a very nice agreement in orientation, shape and substructure of the real cluster and the reconstruction. See Fig.~\ref{simulation_rec}, where we show the convergence map of the reconstruction which uses only four critical-curve points. In addition we performed a more quantiative analysis, by reconstructing the total mass within a certain radius. Here we used the average redshift of the sources and assumed that the mean convergence vanishes at the borders of the field to correct the mass-sheet degeneracy. The result is shown in Fig.~\ref{simulation_comp}, and shows a good agreement with the real cluster mass on all scales. Note that we plot both reconstructions here, which are practically indistinguishable.

\begin{figure}[ht]
\begin{center}
  \includegraphics[width=\hsize]{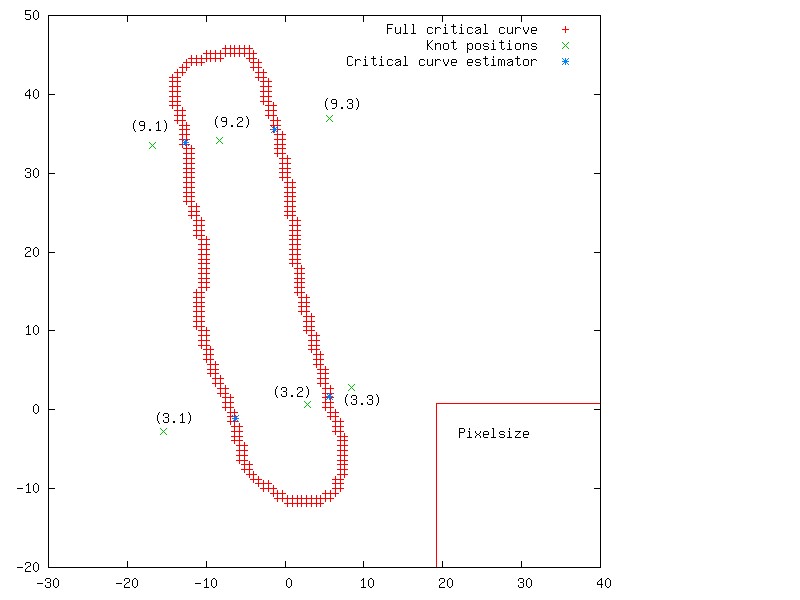}
\end{center}
\caption{This figure illustrates how the critical curve estimators are obtained from given multiple lensed images of the simulation. In the bottom-right corner of the plot we show the pixel size of the final, finely resolved iteration.}
\label{multiple_images}
\end{figure}

\subsection{MS 2137}
\label{ms2137}
We finally apply our reconstruction algorithm to the galaxy cluster MS~2137. It is a cluster at redshift $z_{c}=0.313$, dominated by a bright cD galaxy. It shows a giant tangential arc, a radial arc which was the first to be discovered \citep[see][]{Fort1992}, and three additional arclets. The spectroscopic redshifts of the arcs were determined to be $z_{\text{1}}=1.501$ and $z_{\text{2}}= 1.502$ \citep{Sand2002}. The tangential arc and two of the arclets belong to the same source. The radial arc also produces one counter image. The cluster has been studied several times before, see \citet{Gavazzi2003,Gavazzi2005,Comerford2006} and \citet{Sand2008}. All these studies used parametric reconstruction routines differing from our method. \\

\begin{figure}[ht]
  \includegraphics[width=\hsize]{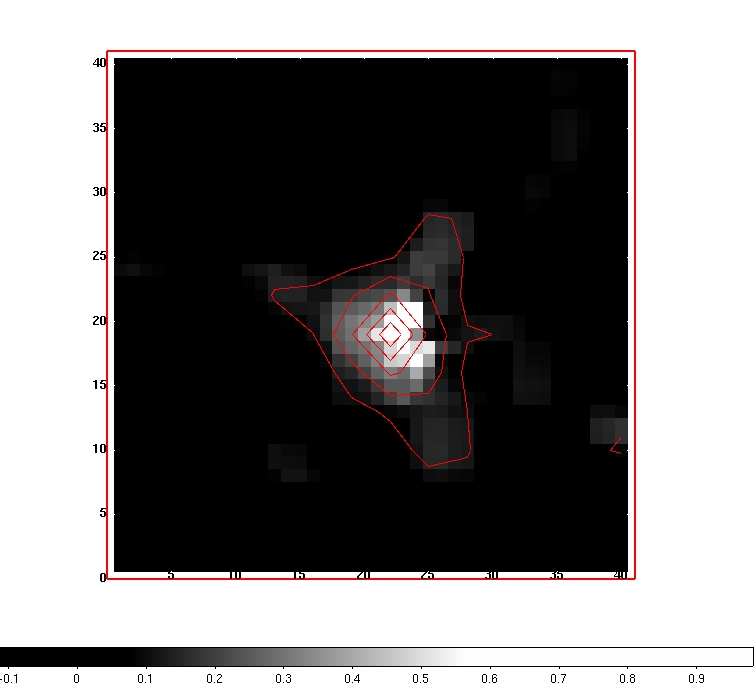}
\caption{High resolution reconstruction of MS~2137 on a refined $40\times40$ pixel grid. The side length corresponds to 1.8~Mpc. The contours start at $\kappa=0.1$ and are spaced by $\Delta\kappa=0.18$}
\label{ms2137_highrec}
\end{figure}

\begin{figure}[ht]
  \includegraphics[width=\hsize]{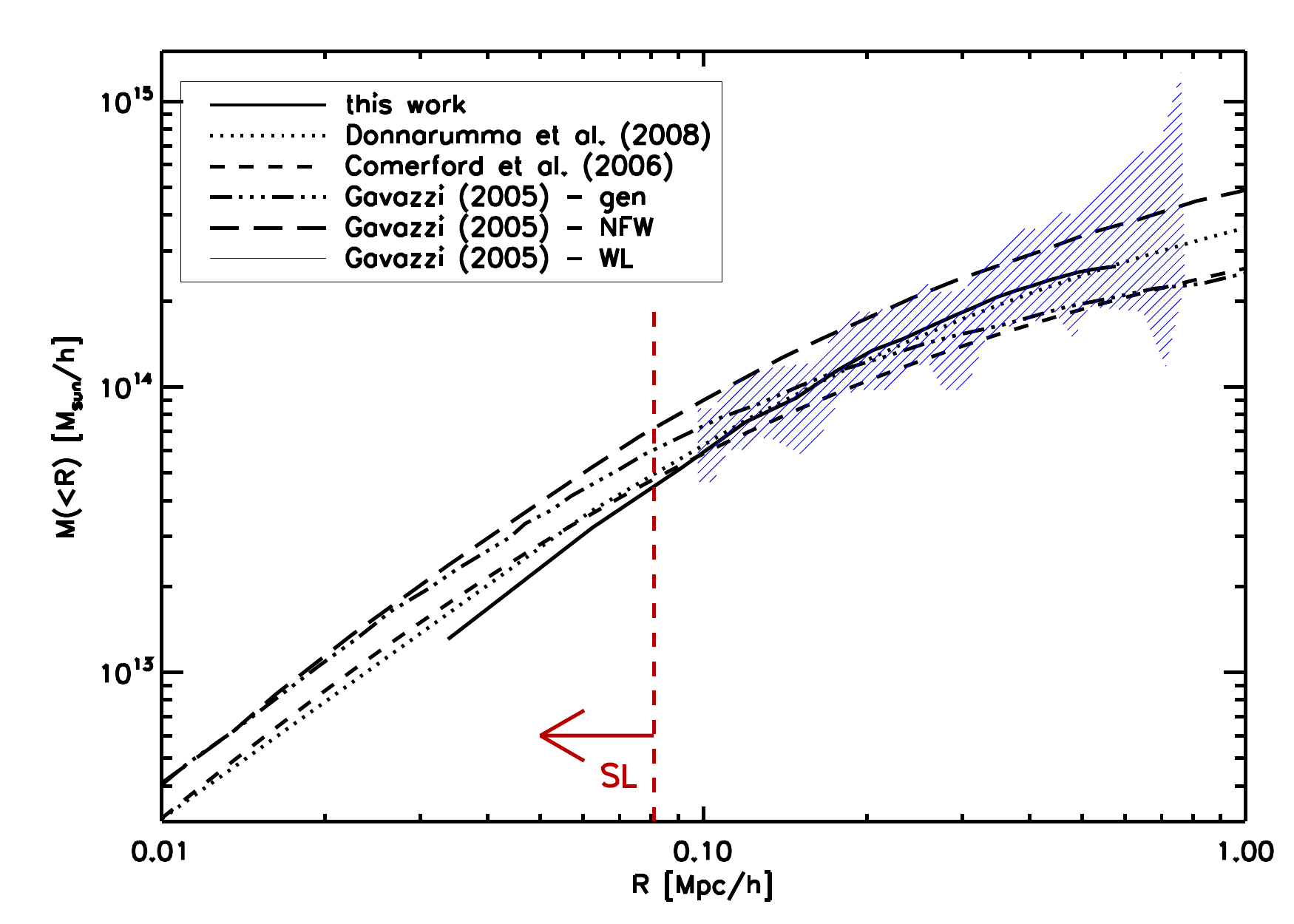}
\caption{Comparison of our results with other reconstructions. The plot shows the reconstructed mass within a certain radius. Also indidcated is the transition between the weak and the strong lensing regime by means of the radius, in which all multiple images of the cluster are contained.}
\label{ms2137_comp}
\end{figure}

The weak-lensing analysis is based on an VLT/FORS observation obtained during August~2001. Its results were kindly provided by R.~Gavazzi (IAP, Paris). The field size is $410^{~\prime\prime}\times410^{~\prime\prime}$. The ellipticity catalogue was obtained with a KSB-method \citep{Kaiser1995} and returned 1500 ellipticity measurements on background galaxies. The arc positions were obtained from an HST/WFPC2 exposure using the F702W filter. During the reconstruction, the measured arc redshifts were used. Based on the experience from the tests with simulated data, we averaged over 15 galaxies per reconstruction pixel. The low-resolution reconstruction was performed on a $25\times25$ pixel grid which was gradually refined to $40\times40$ pixels for a reconstruction which resolves the arcs better (see Fig.~\ref{ms2137_highrec}). Unfortunately, there is a large void in the background-galaxy data in the upper middle part of the field. In this area, the reconstruction is thus unable to resolve any structures.\\
A comparison with former reconstructions is shown in Fig.~\ref{ms2137_comp}. In the weak-lensing regime we have a close agreement with the \citet{Gavazzi2005} results, which is expected, since we are using the same weak lensing data. In the strong-lensing regime we are in good agreement with the latest reconstructions of \citet{Donnarumma2009} and \citet{Comerford2006}, but it tends to prefer a lower central mass than \citet{Gavazzi2005}. The reason for this is still unclear, but the reconstructions of \citet{Comerford2006} and \citet{Donnarumma2009} seem to prefer a lower mass in the strong lensing regime. The discrepancy between those works is discussed in \citet{Donnarumma2009}. 

\section{Conclusions}
\label{conclusions}

We have extended the algorithm proposed by \citet{Cacciato2006} which reconstructs the lensing potential of galaxy clusters on a two-level grid, combining weak and strong-lensing data. Our extensions concern three aspects: First, the shear is replaced by the reduced shear to improve the reconstruction in the mildly non-linear regime. The non-linearity introduced in this way is solved by an inner iteration loop which preserves the linear minimisation of the $\chi^2$ function. Second, the spatial resolution of the potential grid is gradually increased in an outer iteration loop. While this step causes correlations between neighbouring pixels, which need to be dealt with using a non-diagonal covariance matrix, it prepares for the introduction of a highly resolved grid covering the strong-lensing observations in the cluster core. Third, we add a regularisation term to the $\chi^2$ function, avoiding overfitting of noise and allowing a smooth transition from the inner, high-resolution to the outer, coarse-resolution grid. \\
These extensions of the algorithm, especially the introduction of the full $\chi^{2}$-function and the increased number of iterations made it necessary to drastically increase the numerical performance of the reconstruction routines. This was done with fast, numerical multiplication schemes and the parallelisation of the code to run on MPI machines. 
Applications of this algorithm to synthetic data show that the known, simulated cluster lenses are accurately reproduced, although intrinsic ellipticities, measurement and shot noise inevitably lead to a considerable noise level of the reconstruction in the case of realistic data. Still, the quantitative analysis of our convergence maps shows a very good agreement with the expected result. An application to the galaxy cluster MS~2137 qualitatively confirms the results of earlier, parametric studies that show an almost axially symmetric, presumably relaxed mass distribution.
Although there is still room for potentially important further improvements, our algorithm should now be ready for reliable recoveries of lensing potentials, density distributions and density profiles of real galaxy clusters.

\begin{acknowledgements}
This work was supported in part by the Deutsche Forschungsgemeinschaft through the Collaborative Research Centre SFB 439, ``Galaxies in the Young Universe''. J.M. is supported by the Heidelberg Graduate School of Fundamental Physics (HGSFP) and C.M. is supported by the International Max Planck Research School (IMPRS) for Astronomy and Cosmic Physics at the University of Heidelberg. J.M. and M.C. want to thank G. Mamon and R. Gavazzi for useful discussions at IAP. Furthermore, we want to thank F. Bellagamba, M. Radovich and K. Dolag for their contribution to this work.
\end{acknowledgements}


\appendix

\section{Linearisation in grid space}
\label{linearisationingridspace}

The matrix representation of the finite differences (Eqs.~\ref{Kij}, \ref{G1ij} and \ref{G2ij}) allows a simplification of the $\chi^2$ minimisation. For weak lensing, it reads
\begin{equation}
  \chi^{2}_{\text{w}}=\chi^{2}_{1}+\chi^{2}_{2}\;,
\end{equation}
containing one term each for the two ellipticity components.

Of course, we have to perform the minimisation for both components, but we show only the calculation for one component. Correspondingly, the quantities $\varepsilon,\gamma,\mathcal{C},\mathcal{G}$ represent $\varepsilon^{1},\gamma^{1},\mathcal{C}^{1},\mathcal{G}^{1}$ and $\varepsilon^{2},\gamma^{2},\mathcal{C}^{2},\mathcal{G}^{2}$, respectively. One $\chi^2$ contribution is
\begin{equation}
\begin{split}
  \chi^{2}&=
  \left(\varepsilon_{i}-\frac{Z_{i}\gamma_{i}}{1-Z_{i}\kappa_{i}}\right)
  \mathcal{C}^{-1}_{ij}
  \left(\varepsilon_{j}-\frac{Z_{j}\gamma_{j}}{1-Z_{j}\kappa_{j}}\right) \\
  &=\underbrace
   {\frac{\mathcal{C}^{-1}_{ij}}{(1-Z_{i}\kappa_{i})(1-Z_{j}\kappa_{j})}}_{\mathcal{F}_{ij}}
  \left(\varepsilon_{i}(1-Z_{i}\kappa_{i})-Z_{i}\gamma_{i}\right)
  \left(\varepsilon_{j}(1-Z_{j}\kappa_{j})-Z_{j}\gamma_{j}\right) \\
  &=\mathcal{F}_{ij}\left[
    (\varepsilon_{i}-\varepsilon_{i}Z_{i}\kappa_{i}-Z_{i}\gamma_{i})
    (\varepsilon_{j}-\varepsilon_{j}Z_{j}\kappa_{j}-Z_{j}\gamma_{j})
  \right] \\
  &=\mathcal{F}_{ij}\left[
    \varepsilon_{i}\varepsilon_{j}-\varepsilon_{i}\varepsilon_{j}Z_{j}\kappa_{j}-
    \varepsilon_{i}Z_{j}\gamma_{j}-\varepsilon_{i}\varepsilon_{j}Z_{i}\kappa_{i}+
    \varepsilon_{i}\varepsilon_{j}Z_{i}Z_{j}\kappa_{i}\kappa_{j}
  \right.\\
  &\enspace\enspace\enspace\enspace\enspace\left.
    +\varepsilon_{i}Z_{i}Z_{j}\kappa_{i}\gamma_{j}-\varepsilon_{j}Z_{i}\gamma_{i}+
    \varepsilon_{j}Z_{i}Z_{j}\kappa_{j}\gamma_{i}+Z_{i}Z_{j}\gamma_{i}\gamma_{j}
  \right]\;,\\
\end{split}
\end{equation}
where we combined the non-linear factors $(1-Z\kappa)$ in the matrix prefactor $\mathcal{F}_{ij}$. We deal with this term by the iterative approach described in Sect.~\ref{innerleveliteration}. In each iteration, this term is simply kept constant.

Now, we minimise this equation with respect to the potential values $\psi_{l}$,
\begin{equation}
  \frac{\partial\chi^{2}(\psi)}{\partial\psi_{l}}\stackrel{\text{!}}{=}0
\label{minimum}
\end{equation}
The derivative of $\chi^2$ with respect to $\psi_l$ is
\begin{equation}
\begin{split}
  \frac{\partial\chi^{2}(\psi)}{\partial\psi_{l}}=\mathcal{F}_{ij}&\left[
    -\varepsilon_{i}\varepsilon_{j}Z_{j}\frac{\partial}{\partial\psi_{l}}\kappa_{j}(\psi)
    -\varepsilon_{i}Z_{j}\frac{\partial}{\partial\psi_{l}}\gamma_{j}(\psi)
  \right.\\
  &\left.
    -\varepsilon_{i}\varepsilon_{j}Z_{i}\frac{\partial}{\partial\psi_{l}}\kappa_{i}(\psi)
  \right.\\
  &+\varepsilon_{i}\varepsilon_{j}Z_{i}Z_{j}\kappa_{i}
    \frac{\partial}{\partial\psi_{l}}\kappa_{j}(\psi) 
    +\varepsilon_{i}\varepsilon_{j}Z_{i}Z_{j}\kappa_{j}
    \frac{\partial}{\partial\psi_{l}}\kappa_{i}(\psi) \\
  &+\varepsilon_{i}Z_{i}Z_{j}\kappa_{i}\frac{\partial}{\partial\psi_{l}}\gamma_{j}(\psi) 
    +\varepsilon_{i}Z_{i}Z_{j}\gamma_{j}\frac{\partial}{\partial\psi_{l}}\kappa_{i}(\psi) \\
  &-\varepsilon_{j}Z_{i}Z_{j}\frac{\partial}{\partial\psi_{l}}\gamma_{i}(\psi)
    +\varepsilon_{j}Z_{i}Z_{j}\kappa_{j}\frac{\partial}{\partial\psi_{l}}\gamma_{i}(\psi) \\
  &\left.
    +\varepsilon_{j}Z_{i}Z_{j}\gamma_{i}
    \frac{\partial}{\partial\psi_{l}}\kappa_{j}(\psi)+
    Z_{i}Z_{j}\gamma_{i}\frac{\partial}{\partial\psi_{l}}\gamma_{j}(\psi)
  \right. \\
  &\left.
    +Z_{i}Z_{j}\gamma_{j}\frac{\partial}{\partial\psi_{l}}\gamma_{i}(\psi)
  \right]\;.
\end{split}
\end{equation}

Using $\gamma_{i}=\mathcal{G}_{ik}\psi_{k}$, $\kappa_{i}=\mathcal{K}_{ik}\psi_{k}$ and $\frac{\partial}{\partial\psi_{l}}\mathcal{K}_{ik}\psi_{k}=\mathcal{K}_{ik}\delta_{kl}$, we can replace the remaining derivatives,
\begin{equation}
\begin{split}
  \frac{\partial\chi^{2}(\psi_{k})}{\partial\psi_{l}}=\mathcal{F}_{ij}&\left[
    -\varepsilon_{i}\varepsilon_{j}Z_{j}\mathcal{K}_{jk}\delta_{kl}
    -\varepsilon_{i}Z_{j}\mathcal{G}_{jk}\delta_{kl}
    -\varepsilon_{i}\varepsilon_{j}Z_{i}\mathcal{K}_{ik}\delta_{kl}
  \right.\\
  &+\varepsilon_{i}\varepsilon_{j}Z_{i}Z_{j}\mathcal{K}_{ik}\psi_{k}\mathcal{K}_{jk}\delta_{kl} + 
    +\varepsilon_{i}\varepsilon_{j}Z_{i}Z_{j}\mathcal{K}_{jk}\psi_{k}\mathcal{K}_{ik}\delta_{kl}\\
  &+\varepsilon_{i}Z_{i}Z_{j}\mathcal{K}_{ik}\psi_{k}\mathcal{G}_{jk}\delta_{kl} 
    +\varepsilon_{i}Z_{i}Z_{j}\mathcal{G}_{jk}\psi_{k}\mathcal{K}_{ik}\delta_{kl} \\
  &-\varepsilon_{j}Z_{i}Z_{j}\mathcal{G}_{ik}\delta_{kl}
    +\varepsilon_{j}Z_{i}Z_{j}\mathcal{K}_{jk}\psi_{k}\mathcal{G}_{ik}\delta_{kl} \\
  &\left.
    +\varepsilon_{j}Z_{i}Z_{j}\mathcal{G}_{ik}\psi_{k}\mathcal{K}_{jk}\delta_{kl}
    +Z_{i}Z_{j}\mathcal{G}_{ik}\psi_{k}\mathcal{G}_{jk}\delta_{kl}
  \right.\\
  &\left.+Z_{i}Z_{j}\mathcal{G}_{jk}\psi_{k}\mathcal{G}_{ik}\delta_{kl} \right]\;.
\end{split}
\end{equation}

The last equation shows that we can now write Eq.~\ref{minimum} as a linear system of equations,
\begin{equation}
  \mathcal{B}_{lk}\psi_{k}=\mathcal{V}_{l}\;,
\end{equation}
with the coefficient matrix
\begin{equation}
\begin{split}
  \mathcal{B}_{lk}=\mathcal{F}_{ij}&\left[
    \varepsilon_{i}\varepsilon_{j}Z_{i}Z_{j}\mathcal{K}_{ik}\mathcal{K}_{jl}+
    \varepsilon_{i}\varepsilon_{j}Z_{i}Z_{j}\mathcal{K}_{jk}\mathcal{K}_{il}+
    \varepsilon_{i}Z_{i}Z_{j}\mathcal{K}_{ik}\mathcal{G}_{jl}
  \right. \\
  &+\varepsilon_{i}Z_{i}Z_{j}\mathcal{G}_{jk}\mathcal{K}_{il}
    +\varepsilon_{j}Z_{i}Z_{j}\mathcal{K}_{jk}\mathcal{G}_{il}
    +\varepsilon_{i}Z_{i}Z_{j}\mathcal{G}_{ik}\mathcal{K}_{jl} \\
  &\left.
    +Z_{i}Z_{j}\mathcal{G}_{ik}\mathcal{G}_{jl}+Z_{i}Z_{j}\mathcal{G}_{jk}\mathcal{G}_{il}
  \right]\;,
\end{split}
\end{equation}
and the result vector
\begin{equation}
  \mathcal{V}_{l}=\mathcal{F}_{ij}\left[
    \varepsilon_{i}\varepsilon_{j}\mathcal{K}_{jl}+
    \varepsilon_{i}Z_{j}\mathcal{G}_{jl}+
    \varepsilon_{i}\varepsilon_{j}Z_{i}\mathcal{K}_{il}+
    \varepsilon_{j}Z_{i}\mathcal{G}_{il}
  \right]\;.
\end{equation}

We now repeat this exercise for the strong lensing term, where
\begin{equation}
  \chi^{2}_{\text{s}}=\frac{(\det\mathcal{A})^{2}_{i}}{\sigma^{2}_{i}}=
  \frac{((1-Z_{i}\kappa_{i})^{2}-|Z_{i}\gamma_{i}|^{2})^{2}}{\sigma^{2}_{i}}\;.
\end{equation}
Again, we isolate the non-linear terms and take them as constant during each iteration step,
\begin{equation}
\begin{split}
  &\frac{\partial\chi^{2}_{\text{s}}(\psi_{k})}{\partial\psi_{l}}=
  \frac{2(\det\mathcal{A})_{i}}{\sigma^{2}_{i}}
  \frac{\partial}{\partial\psi_{l}}(\det\mathcal{A}(\psi_{k}))_{i} \\
  &=\frac{2(\det\mathcal{A})_{i}}{\sigma^{2}_{i}}\left[
    \frac{\partial}{\partial\psi_{l}}(1-Z_{i}\kappa_{i}(\psi_{k})^{2}-
    \frac{\partial}{\psi_{l}}|Z_{i}\gamma_{i}(\psi_{k})|^{2}
  \right] \\
  &=\frac{2(\det\mathcal{A})_{i}}{\sigma^{2}_{i}}\left[
    2(1-Z_{i}\kappa_{i}(\psi_{k}))\left(
      -Z_{i}\frac{\partial}{\partial\psi_{l}}\kappa_{i}(\psi_{k})
    \right) \right.\\
  &\left.-\frac{\partial}{\partial\psi_{l}}\left(
    Z^{2}_{i}\gamma^{2}_{1i}(\psi_{k})+Z^{2}_{i}\gamma^{2}_{2i}(\psi_{k})
  \right)\right] \\
  &=\frac{2(\det\mathcal{A})_{i}}{\sigma^{2}_{i}}\left[
    2(1-Z_{i}\kappa_{i}(\psi_{k}))(-Z_{i}\mathcal{K}_{il})-
    2Z_{i}\gamma_{1i}(\psi_{k})Z_{i}\mathcal{G}^{1}_{il}
  \right.\\
  &\left.
    -2Z_{i}\gamma_{2i}(\psi_{k})Z_{i}\mathcal{G}^{2}_{il}
  \right] \\
  &=\frac{4(\det\mathcal{A})_{i}}{\sigma^{2}_{i}}\left[
    Z_{i}^{2}\mathcal{K}_{ik}\mathcal{K}_{il}\psi_{k}-
    Z_{i}\mathcal{K}_{il}-Z^{2}_{i}\mathcal{G}^{1}_{ik}\mathcal{G}^{1}_{il}\psi_{k}-
    Z_{i}\mathcal{K}_{il}
  \right.\\
  &\left.-Z^{2}_{i}\mathcal{G}^{2}_{ik}\mathcal{G}^{2}_{il}\psi_{k}\right]\;.
\end{split}
\end{equation}

This yields another linear system,
\begin{equation}
  \mathcal{B}^{*}_{lk}\psi_{k}=\mathcal{V}^{*}_{l}\;,
\end{equation}
with the coefficient matrix for strong lensing
\begin{equation}
  \mathcal{B}^{*}_{lk}=\frac{4(\det\mathcal{A})_{i}}{\sigma^{2}_{i}}
  Z^{2}_{i}(\mathcal{K}_{ik}\mathcal{K}_{il}-
  \mathcal{G}^{1}_{ik}\mathcal{G}^{1}_{il}-
  \mathcal{G}^{2}_{ik}\mathcal{G}^{2}_{il})
\end{equation}
and the result vector
\begin{equation}
  \mathcal{V}^{*}_{l}=\frac{4(\det\mathcal{A})_{i}}{\sigma^{2}_{i}}Z_{i}\mathcal{K}_{il}\;.
\end{equation}

Also the regularisation term in Eq.~(\ref{R}) has to be minimised,
\begin{equation}
\begin{split}
  \frac{\partial R(\psi_{k})}{\partial\psi_{l}}
  &=\eta_{i}\frac{\partial}{\partial\psi_{l}}\left(\kappa^{b}_{i}
    -\kappa_{i}(\psi_{k})\right)^{2} \\
  &=2\eta_{i}(\kappa^{b}_{i}-\kappa_{i})\left(
    -\frac{\partial}{\partial\psi_{l}}\mathcal{K}_{ik}\psi_{k}
  \right) \\
  &=2\eta_{i}\left(\kappa_{i}^{b}-\mathcal{K}_{ik}\psi_{k}\right)
    \left(-\mathcal{K}_{il}\right) \\
  &=2\eta_{i}\left(
    -\kappa^{b}_{i}\mathcal{K}_{il}+\mathcal{K}_{ik}\mathcal{K}_{il}\psi_{k}
  \right)\;,
\end{split}
\end{equation}
which contributes one additional term to the coefficient matrix and the result vector,
\begin{equation}
  \mathcal{B}^{\text{reg}}_{lk}=\sum\limits_{i}\eta_{i}\mathcal{K}_{ik}\mathcal{K}_{il}
\end{equation}
and
\begin{equation}
  \mathcal{V}^{\text{reg}}_{l}=\sum\limits_{i}\eta_{i}\kappa^{b}_{i}\mathcal{K}_{il}\;.
\end{equation}

Finally, we collect the results to obtain the solution for Eq.~\ref{chimin}, given in terms of a linear system with the following coefficient matrix
\begin{equation}
\label{blkvl}
\begin{split}
  \mathcal{B}_{lk}=&\sum\limits_{i,j}\mathcal{F}^{1}_{ij}Z_{i}Z_{j} \\
  &\cdot\left[
    \varepsilon^{1}_{i}\varepsilon^{1}_{j}\mathcal{K}_{ik}\mathcal{K}_{jl}+
    \varepsilon^{1}_{i}\varepsilon^{1}_{j}\mathcal{K}_{jk}\mathcal{K}_{il}+
    \varepsilon^{1}_{i}\mathcal{K}_{ik}\mathcal{G}^{1}_{jl}+
    \varepsilon^{1}_{i}\mathcal{G}^{1}_{jk}\mathcal{K}_{il}
  \right.\\
  &\left.
    +\varepsilon^{1}_{j}\mathcal{K}_{jk}\mathcal{G}^{1}_{il}
    +\varepsilon^{1}_{j}\mathcal{G}^{1}_{ik}\mathcal{K}_{jl}
    +\mathcal{G}^{1}_{ik}\mathcal{G}^{1}_{jl}
    +\mathcal{G}^{1}_{jk}\mathcal{G}^{1}_{il}
  \right] \\
  &+\sum\limits_{i,j}\mathcal{F}^{2}_{ij}Z_{i}Z_{j} \\
  &\cdot\left[
    \varepsilon^{2}_{i}\varepsilon^{2}_{j}\mathcal{K}_{ik}\mathcal{K}_{jl}+
    \varepsilon^{2}_{i}\varepsilon^{2}_{j}\mathcal{K}_{jk}\mathcal{K}_{il}+
    \varepsilon^{2}_{i}\mathcal{K}_{ik}\mathcal{G}^{2}_{jl}+
    \varepsilon^{2}_{i}\mathcal{G}^{2}_{jk}\mathcal{K}_{il}
  \right.\\
  &\left.
    +\varepsilon^{2}_{j}\mathcal{K}_{jk}\mathcal{G}^{2}_{il}
    +\varepsilon^{2}_{j}\mathcal{G}^{2}_{ik}\mathcal{K}_{jl}
    +\mathcal{G}^{2}_{ik}\mathcal{G}^{2}_{jl}
    +\mathcal{G}^{2}_{jk}\mathcal{G}^{2}_{il}
  \right] \\
  &+\sum\limits_{m}\frac{4(\det\mathcal{A})_{m}}{\sigma^{2}_{m}}Z^{2}_{m}\\
  &\cdot\left[
    \mathcal{K}_{mk}\mathcal{K}_{ml}-\mathcal{G}^{1}_{mk}\mathcal{G}^{1}_{ml}-
    \mathcal{G}^{2}_{mk}\mathcal{G}^{2}_{ml}
  \right] \\
  &+\sum\limits_{n}\eta_{n}\mathcal{K}_{nk}\mathcal{K}_{nl}
\end{split}
\end{equation}
and the result vector
\begin{equation}
\begin{split}
  \mathcal{V}_{l}&=\sum\limits_{i,j}\mathcal{F}^{1}_{ij}\left[
    \varepsilon^{1}_{i}\varepsilon^{1}_{j}\mathcal{K}_{jl}+
    \varepsilon^{1}_{i}Z_{j}\mathcal{G}^{1}_{jl}+
    \varepsilon^{1}_{i}\varepsilon^{1}_{j}Z_{i}\mathcal{K}_{il}+
    \varepsilon^{1}_{j}Z_{i}\mathcal{G}^{1}_{il}
  \right] \\
  &+\sum\limits_{i,j}\mathcal{F}^{2}_{ij}\left[
    \varepsilon^{2}_{i}\varepsilon^{2}_{j}\mathcal{K}_{jl}+
    \varepsilon^{2}_{i}Z_{j}\mathcal{G}^{2}_{jl}+
    \varepsilon^{2}_{i}\varepsilon^{2}_{j}Z_{i}\mathcal{K}_{il}+
    \varepsilon^{2}_{j}Z_{i}\mathcal{G}^{2}_{il}
  \right] \\
  &+\sum\limits_{m}
  \frac{4(\det\mathcal{A})_{m}}{\sigma^{2}_{m}}Z_{m}\mathcal{K}_{ml} \\
  &+\sum\limits_{n}\eta_{n}\kappa^{b}_{n}\mathcal{K}_{nl}\;,
\end{split}
\end{equation}
where $i, j, n$ indicate the summation over the complete grid, and $m$ over those pixels which are assumed to be traversed by a critical curve.

\bibliographystyle{aa}
\bibliography{bibliography.bib}

\begin{thebibliography}{28}
\expandafter\ifx\csname natexlab\endcsname\relax\def\natexlab#1{#1}\fi

\bibitem[{{Bartelmann} {et~al.}(1998){Bartelmann}, {Huss}, {Colberg},
  {Jenkins}, \& {Pearce}}]{Bartelmann1998}
{Bartelmann}, M., {Huss}, A., {Colberg}, J.~M., {Jenkins}, A., \& {Pearce},
  F.~R. 1998, \aap, 330, 1

\bibitem[{{Bartelmann} {et~al.}(1996){Bartelmann}, {Narayan}, {Seitz}, \&
  {Schneider}}]{Bartelmann1996}
{Bartelmann}, M., {Narayan}, R., {Seitz}, S., \& {Schneider}, P. 1996, \apjl,
  464, L115+

\bibitem[{{Bartelmann} \& {Schneider}(2001)}]{Bartelmann2001}
{Bartelmann}, M. \& {Schneider}, P. 2001, \physrep, 340, 291

\bibitem[{{Brada{\v c}} {et~al.}(2004){Brada{\v c}}, {Lombardi}, \&
  {Schneider}}]{Bradav2004}
{Brada{\v c}}, M., {Lombardi}, M., \& {Schneider}, P. 2004, \aap, 424, 13

\bibitem[{{Brada{\v c}} {et~al.}(2005){Brada{\v c}}, {Schneider}, {Lombardi},
  \& {Erben}}]{Bradav2005}
{Brada{\v c}}, M., {Schneider}, P., {Lombardi}, M., \& {Erben}, T. 2005, \aap,
  437, 39

\bibitem[{{Cacciato} {et~al.}(2006){Cacciato}, {Bartelmann}, {Meneghetti}, \&
  {Moscardini}}]{Cacciato2006}
{Cacciato}, M., {Bartelmann}, M., {Meneghetti}, M., \& {Moscardini}, L. 2006,
  \aap, 458, 349

\bibitem[{{Coe} {et~al.}(2006){Coe}, {Ben{\'{\i}}tez}, {S{\'a}nchez}, {Jee},
  {Bouwens}, \& {Ford}}]{Coe2006}
{Coe}, D., {Ben{\'{\i}}tez}, N., {S{\'a}nchez}, S.~F., {et~al.} 2006, \aj, 132,
  926

\bibitem[{{Comerford} {et~al.}(2006){Comerford}, {Meneghetti}, {Bartelmann}, \&
  {Schirmer}}]{Comerford2006}
{Comerford}, J.~M., {Meneghetti}, M., {Bartelmann}, M., \& {Schirmer}, M. 2006,
  \apj, 642, 39

\bibitem[{{Diego} {et~al.}(2007){Diego}, {Tegmark}, {Protopapas}, \&
  {Sandvik}}]{Diego2007}
{Diego}, J.~M., {Tegmark}, M., {Protopapas}, P., \& {Sandvik}, H.~B. 2007,
  \mnras, 375, 958

\bibitem[{{Dolag} {et~al.}(2005){Dolag}, {Vazza}, {Brunetti}, \&
  {Tormen}}]{Dolag2005}
{Dolag}, K., {Vazza}, F., {Brunetti}, G., \& {Tormen}, G. 2005, \mnras, 364,
  753

\bibitem[{{Donnarumma} {et~al.}(2009){Donnarumma}, {Ettori}, {Meneghetti}, \&
  {Moscardini}}]{Donnarumma2009}
{Donnarumma}, A., {Ettori}, S., {Meneghetti}, M., \& {Moscardini}, L. 2009,
  ArXiv e-prints, astro-ph/0902.4051

\bibitem[{{Falco} {et~al.}(1985){Falco}, {Gorenstein}, \&
  {Shapiro}}]{Falco1985}
{Falco}, E.~E., {Gorenstein}, M.~V., \& {Shapiro}, I.~I. 1985, \apjl, 289, L1

\bibitem[{{Fort} {et~al.}(1992){Fort}, {Le Fevre}, {Hammer}, \&
  {Cailloux}}]{Fort1992}
{Fort}, B., {Le Fevre}, O., {Hammer}, F., \& {Cailloux}, M. 1992, \apjl, 399,
  L125

\bibitem[{{Gavazzi}(2005)}]{Gavazzi2005}
{Gavazzi}, R. 2005, \aap, 443, 793

\bibitem[{{Gavazzi} {et~al.}(2003){Gavazzi}, {Fort}, {Mellier}, {Pell{\'o}}, \&
  {Dantel-Fort}}]{Gavazzi2003}
{Gavazzi}, R., {Fort}, B., {Mellier}, Y., {Pell{\'o}}, R., \& {Dantel-Fort}, M.
  2003, \aap, 403, 11

\bibitem[{{Kaiser} {et~al.}(1995){Kaiser}, {Squires}, \&
  {Broadhurst}}]{Kaiser1995}
{Kaiser}, N., {Squires}, G., \& {Broadhurst}, T. 1995, \apj, 449, 460

\bibitem[{{Madau} {et~al.}(2008){Madau}, {Diemand}, \& {Kuhlen}}]{Madau2008}
{Madau}, P., {Diemand}, J., \& {Kuhlen}, M. 2008, \apj, 679, 1260

\bibitem[{{Meneghetti} {et~al.}(2007){Meneghetti}, {Argazzi}, {Pace},
  {Moscardini}, {Dolag}, {Bartelmann}, {Li}, \& {Oguri}}]{Meneghetti2007a}
{Meneghetti}, M., {Argazzi}, R., {Pace}, F., {et~al.} 2007, \aap, 461, 25

\bibitem[{{Meneghetti} {et~al.}(2008){Meneghetti}, {Melchior}, {Grazian}, {De
  Lucia}, {Dolag}, {Bartelmann}, {Heymans}, {Moscardini}, \&
  {Radovich}}]{Meneghetti2008}
{Meneghetti}, M., {Melchior}, P., {Grazian}, A., {et~al.} 2008, \aap, 482, 403

\bibitem[{{Narayan} \& {Bartelmann}(1996)}]{Narayan1996}
{Narayan}, R. \& {Bartelmann}, M. 1996, ArXiv Astrophysics e-prints,
  astro-ph/9606001

\bibitem[{{Navarro} {et~al.}(1996){Navarro}, {Frenk}, \& {White}}]{Navarro1996}
{Navarro}, J.~F., {Frenk}, C.~S., \& {White}, S.~D.~M. 1996, \apj, 462, 563

\bibitem[{P.~Schneider(1992)}]{P.1992}
P.~Schneider, J.~Ehlers, E.~F. 1992, Gravitational Lenses (Springer Verlag)

\bibitem[{P.~Schneider(2006)}]{P.2006}
P.~Schneider, C.~Kochanek, J.~W. 2006, Gravitational Lensing: Strong, Weak and
  Micro (Springer Verlag)

\bibitem[{{Puchwein} {et~al.}(2005){Puchwein}, {Bartelmann}, {Dolag}, \&
  {Meneghetti}}]{Puchwein2005}
{Puchwein}, E., {Bartelmann}, M., {Dolag}, K., \& {Meneghetti}, M. 2005, \aap,
  442, 405

\bibitem[{{Sand} {et~al.}(2002){Sand}, {Treu}, \& {Ellis}}]{Sand2002}
{Sand}, D.~J., {Treu}, T., \& {Ellis}, R.~S. 2002, \apjl, 574, L129

\bibitem[{{Sand} {et~al.}(2008){Sand}, {Treu}, {Ellis}, {Smith}, \&
  {Kneib}}]{Sand2008}
{Sand}, D.~J., {Treu}, T., {Ellis}, R.~S., {Smith}, G.~P., \& {Kneib}, J.-P.
  2008, \apj, 674, 711

\bibitem[{{Seitz} {et~al.}(1998){Seitz}, {Schneider}, \&
  {Bartelmann}}]{Seitz1998}
{Seitz}, S., {Schneider}, P., \& {Bartelmann}, M. 1998, \aap, 337, 325

\bibitem[{{Springel} {et~al.}(2008){Springel}, {Wang}, {Vogelsberger},
  {Ludlow}, {Jenkins}, {Helmi}, {Navarro}, {Frenk}, \& {White}}]{Springel2008}
{Springel}, V., {Wang}, J., {Vogelsberger}, M., {et~al.} 2008, \mnras, 391,
  1685

\end{thebibliography}

\end{document}